\documentclass[10pt,journal,compsoc]{IEEEtran}

\ifCLASSOPTIONcompsoc
  \usepackage[nocompress]{cite}
\else
  \usepackage{cite}
\fi

\usepackage{amsmath,amsfonts}
\usepackage{algorithmic}
\usepackage{algorithm}

\usepackage{array}
\usepackage{textcomp}
\usepackage{stfloats}
\usepackage{url}
\usepackage{verbatim}
\usepackage{graphicx}
\usepackage{cite}
\usepackage{enumitem}
\usepackage{subfigure}
\usepackage{multirow}
\usepackage{booktabs}
\usepackage{textcomp}
\usepackage{xcolor}

\usepackage{enumerate}

\begin{document}

\title{Cross-domain Transfer of Valence Preferences via a Meta-optimization Approach}

\author{Chuang~Zhao,
        Hongke~Zhao, Ming~He, Xiaomeng~Li~\IEEEmembership{Member, IEEE}, Jianping~Fan
        
\IEEEcompsocitemizethanks{
\IEEEcompsocthanksitem C. Zhao is with the Department of Electronic and Computer Engineering, The Hong Kong University of Science and Technology, Hong Kong, SAR, China;  the College of Management and Economics, Tianjin University, Tianjin 30072, China (e-mail:
zhaochuang\_01@163.com). This work was done when C. Zhao was doing an internship at AI Lab of Lenovo Research. 

\IEEEcompsocthanksitem H. Zhao is with the College of Management and Economics, Laboratory of Computation and Analytics of Complex Management Systems (CACMS), Tianjin University, Tianjin 30072, China; (e-mail: hongke@tju.edu.cn) H. Zhao is the corresponding author.

\IEEEcompsocthanksitem X. Li is with the Department of Electronic and Computer Engineering, The Hong Kong University of Science and Technology, Hong Kong, SAR, China, and also with The Hong Kong University of Science and Technology Shenzhen Research Institute, Shenzhen 518057, China (e-mail: eexmli@ust.hk) 

\IEEEcompsocthanksitem M. He and J. Fan are with AI Lab at Lenovo Research, Beijing 100000, China (e-mail: heming01@foxmail.com; jfan1@lenovo.com;).
}}

\markboth{IEEE Transactions on Pattern Analysis and Machine Intelligence
}%
{Shell \MakeLowercase{\textit{et al.}}: A Sample Article Using IEEEtran.cls for IEEE Journals}

\IEEEtitleabstractindextext{%
\begin{abstract}
Cross-domain recommendation offers a potential avenue for alleviating data sparsity and cold-start problems by transferring knowledge from an auxiliary domain to the target domain.
Embedding and mapping, as a classic cross-domain research genre, aims to identify a common mapping function to perform representation transformation between two domains via exploiting the supervision signals of overlapping users.
Nevertheless, previous coarse-grained preference representations, non-personalized mapping functions, and excessive reliance on overlapping users limit their performance, especially in scenarios where overlapping users are sparse.
To address aforementioned challenges, we propose a novel \textit{\textbf{C}ross-domain transfer of \textbf{V}alence \textbf{P}references via a \textbf{M}eta-optimization approach}, namely \textbf{\textit{CVPM}}. 
\textit{CVPM} formalizes cross-domain interest transfer as a hybrid architecture of parametric meta-learning and self-supervised learning, which not only transfers user preferences at a finer level, but also enables signal enhancement with the knowledge of non-overlapping users.
Specifically, with deep insights into user preferences and valence preference theory, we believe that there exists significant difference between users' positive preferences and negative behaviors, and thus employ differentiated encoders to learn their distributions. In particular, we further utilize the pre-trained model and item popularity to sample pseudo-interaction items to ensure the integrity of both distributions.
To guarantee the personalization of preference transfer, we treat each user's mapping as two parts, the common transformation and the personalized bias, where the network used to generate the personalized bias is output by a meta-learner.
Furthermore, in addition to the supervised loss for overlapping users, we design contrastive tasks for non-overlapping users from both group and individual-levels to avoid model skew and enhance the semantics of representations.
We construct 5 cross-domain tasks and 1 cross-system task from 8 data sets, and evaluate the model performance under two scenarios of cold start and warm start. Exhaustive data analysis and extensive experimental results demonstrate the effectiveness and advancement of our proposed framework.
\end{abstract}

}

\maketitle

\IEEEdisplaynontitleabstractindextext

%
\IEEEpeerreviewmaketitle

\IEEEraisesectionheading{\section{Introduction}\label{sec:introduction}}

\IEEEPARstart{R}{ecommender} systems~\cite{zhang2023personalized,wang2022disentangled,zhangjietpaimi} are developed to conquer the increasingly serious information overload, a phenomenon in which massive choices are harmful to user engagement, and thus play an indispensable role in e-commerce~\cite{sethi2022survey}, games~\cite{hare2022player}, and online services~\cite{liu2023dual}.
Most traditional recommender systems learn collaborative filtering relationships between users and items from massive interaction data to infer user preferences~\cite{wu2022survey}. Clearly, rich historical user-item interactions are crucial for building high-quality collaborative embeddings. Nevertheless, in real scenarios, there are a large number of cold-start users or cold-start items with little or no interaction records, which poses serious challenges to these approaches~\cite{zang2022survey,du2022metakg}. 

One dazzling pearl to tackle this dilemma is the cross-domain recommender system~\cite{zang2022survey,zhao2023cross}, whose essential idea is to leverage one or more associated auxiliary domains to introduce prior knowledge, thereby enhancing the recommendation ability of the target domain.
For instance, a user in Amazon who has few purchase records might have hundreds of clicking interactions in Netflix. Such observed interaction feedback by Netflix could provide a valuable clue to infer her preference and make recommendations in Amazon.
Prevailing cross-domain recommendation works fall into two genres, content-based and transfer-based~\cite{zhu2021cross}, both relying on overlapping entities (users or items) to facilitate knowledge transfer.
The first genre focuses on using more side information, such as tags~\cite{tang2023cross}, semantic knowledge~\cite{liu2022cross}, user content~\cite{liu2022collaborative}, to establish more connections between entities in two domains for knowledge transfer. Despite great performance, application scenarios are limited due to high data quality requirements. In contrast, transfer-based approaches work on scenarios without auxiliary information, which focus
on transferring latent factors or rating patterns~\cite{zhu2022personalized,cao2022cross}. Our research is dedicated to elevating the second research paradigm.

\begin{figure}[!t]
	\centering
	\includegraphics[width=\linewidth,height=0.4\linewidth]{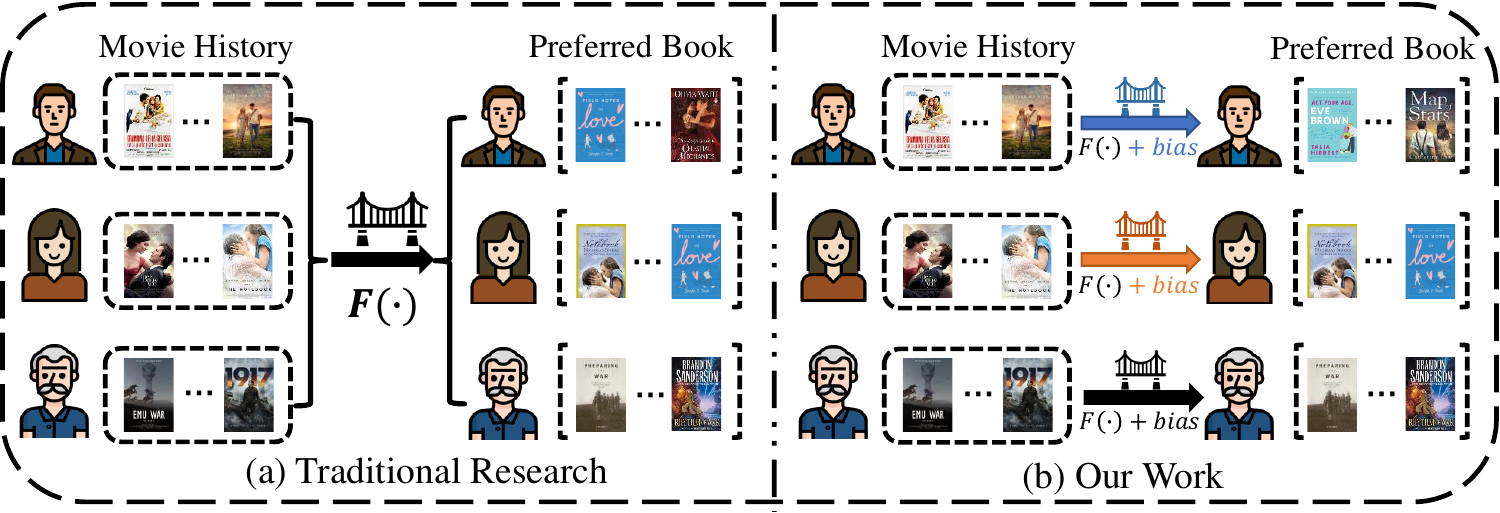}
	\caption{Differences between \textit{CVPM} and previous work. Previous work utilize a common mapping function, while ours is individualized.}
	\label{fig1}
\end{figure}
A common transfer-based approach is to encode users' source and target domain preferences as embeddings, and then explicitly learn a linear or nonlinear common mapping function with overlapping users~\cite{man2017cross,zhu2021transfer}.
On account of this, it is critical to find a reasonable and effective mapping function across domains. Even with remarkable achievements, previous work still suffers from several inherent limitations.
First, heterogeneity in preferences has not been well explored. Existing studies tend to utilize 
historical interactions of users to construct a unified transfer representation~\cite{wang2021low,zhu2022personalized,zhao2021coea}, but abundant evidence~\cite{liu2022review, wang2022invariant,etkin2015neural} indicates that user behavior is driven by diverse preferences and there are noticeable variations in data distribution and rating patterns among different preferences.
Second, the common mapping function  violates the personalized tenet of the recommender system~\cite{guo2020survey,dhelim2022survey}. 
Without a personalized mapping function, recommendations will be dominated by general trends and patterns across domains, and may miss important nuances and individual differences.
In particular, the representation projected by common mapping function may be biased toward users or items with more cross-domain behaviors, regardless of whether it matches the user's personal taste or not. In contrast, through the personalized mapping function, the recommendation can take into account the user's personalized behaviors such as purchases and browsing history in the source domain, resulting in more accurate and targeted personalized transfers that better match the user's preferences.
Third, supervision signals based on overlapping users are prone to skew the models~\cite{yu2021mixed,cao2022cross}. Prevailing approaches have focused on finding a mapping function to shrink the distance between cross-domain representations of overlapping users~\cite{ma2021trust,li2021dual}, thus using data sets that contain a high proportion of overlapping users, or even complete overlap~\cite{chen2023toward}. Nontheless, in the real scenario, the proportion of overlapping users is exceedingly small, even less than 5\%, as evidenced in our data set. This means that the system over-reliance on overlapping users is  insufficient to imply comprehensive collaborative filtering relations and personal tastes, and thus leads to inferior recommendations. 

Motivated by aforementioned issues, in this paper, we propose a \textit{\textbf{C}ross-domain transfer of \textbf{V}alence \textbf{P}references
via a \textbf{M}eta-optimization approach}, namely \textbf{\textit{CVPM}}, which tailors favorable embedding and mapping schemas for effective cross-domain recommendation. \textit{CVPM} is a modern hybrid system of meta-learning and self-supervised learning in pursuit of more accurate and personalized mapping functions.
More precisely, in order to transfer preferences \emph{with greater granularity}, we draw on the valence preference theory~\cite{liu2022review,kato2022rating, kumar2020movie,ullah2016valence} to describe user behavior in the source domain as two distinct distributions, i.e., \textit{positive preference} and \textit{negative preference}. 
Given that the rating patterns and data distributions for the two preferences exhibit significant differences, we employ a decoupled probabilistic learning strategy that can effectively mitigate noise and enhance overall robustness.
For the sake of the \emph{personalized transfer}, our approach goes beyond the typical mapping function by additionally adopting a parameter-based meta-learning framework to create a distinct bias network for each user. This enables our model to leverage both the \textit{global transfer patterns} captured by the common mapping function and the \textit{personalized preferences} of each user, resulting in highly tailored and customized recommendations. Fig.~\ref{fig1} presents the key differences from existing work.
To extract knowledge from non-overlapping users and enhance the semantics of representations, we creatively design contrastive tasks~\cite{lin2022improving,xie2022contrastive}, i.e., distinguishing instances from both \textit{group-level} and \textit{individual-level} perspectives.
Specifically, for the group level, we construct the transferred representations with their assigned cluster centroids in the source domain as positive views, and negative as negative pairs, to compute a contrastive loss. At the individual level, we naturally treat users' negative and positive preferences as corresponding views of the transferred representations.
These two moves suggest that our model not only needs to satisfy the supervision signal of overlapping users, but also ought to consider the instance discriminative properties of non-overlapping users, thus avoiding model skew and missing patterns.
In addition, unlike previous methods that only validate \textit{cold-start cross-domain recommendation scenarios}, we also test \textit{CVPM} in three scenarios: \textit{warm-start cross-domain recommendation}~\footnote{Warm start refers to the situation where the user for testing has limited interaction in the target domain.}, \textit{cold-start cross-system recommendation~\footnote{Cross-system recommendation refers to partial overlap on the item side. For example, Movielens and Netflix partially overlap on movies.}}, and \textit{warm-start cross-system recommendation}. In particular, we fine-tune in the testing phase for warm-start scenarios, and transfer knowledge on the item side for cross-system scenarios.

The main contributions of this work are as follows:
\begin{itemize}[leftmargin=12pt]
    \item To the best of our knowledge, we have made remarkable contributions to cross-domain recommendation by simultaneously considering fine-grained preference characterization, user-specific interest transfer, and non-overlapping user knowledge utilization. These approaches allow user interest transfer with an accurate, comprehensive and personalized manner.
    \item We reconstruct previous embedding and mapping-based cross-domain recommendation into an innovative hybrid architecture of meta-learning and self-supervised learning, which can significantly enhance the semantics of user representations and improve recommendation performance.
    \item We build 5 cross-domain recommendation and 1 cross-system recommendation tasks on 8 data sets, and test the performance of the algorithm in both cold start and warm start scenarios. Extensive experimental results and exhaustive analysis demonstrate the progressiveness of the proposed framework. 
    We have released the sample data set and source code in the \textit{github repository}~\footnote{https://github.com/Data-Designer/MapRec} to ensure its reproducibility and comparability.
\end{itemize}
The rest of the paper is organized as follows. Section~\ref{section2} presents related work, followed by traditional piplines in Section~\ref{section3} and our proposed method in Section~\ref{section4}. Section~\ref{section5} presents our experimental details and analysis, and a detailed discussion is given in Section~\ref{section6}. Finally, we summarize this paper in Section~\ref{section7}.

\section{Related Work}\label{section2}
In this section, we conduct a detailed literature survey on the three research areas most relevant to this work, namely \textit{cross-domain recommendation, meta-learning and self-supervised learning}, and highlight their connections and differences with this study.
\subsection{Cross-domain Recommendation}\label{tax}
Data sparsity and cold-start problems severely restrict the effectiveness of recommender systems, as they require massive data training to capture collaborative filtering relationships~\cite{wu2023cold,guo2020survey}.
Cross-domain recommender systems, which hope to transfer knowledge from rich information source domains, are effective means to get rid of this dilemma~\cite{khan2017cross}.
In terms of mainstream research paradigms, it can be further subdivided into \textit{collective matrix factorization, representation combination of overlapping users, embedding and mapping, and graph neural network-based} approaches~\cite{zang2022survey}.
We are conducting research in the field of embedding and mapping, with the aim of developing a mapping function that can effectively represent cold-start users in the target domain.
EMCDR~\cite{man2017cross} operates under the assumption that a mapping function, either linear or non-linear, can be used to project representations from different domains into a shared space. By leveraging the representations of overlapping users, the mapping function is optimized.
LACDR~\cite{wang2021low} believes that the mapping function suffers from embedding size and missing knowledge, and thus designs a reconstruction loss to enhance representational semantics.
SSCDR~\cite{kang2019semi} performs signal enhancement for non-overlapping users from the perspective of semi-supervised learning, and projects the source domain items through the mapping function to ensure that the triangle relationship remains unchanged.
DCDCSR~\cite{zhu2020deep} argues that the inaccuracy of representation mapping stems from an imbalance in the interaction of the two domains, and thus encourages the target domain representation to be aligned with the benchmark representation built using sparsity.
PTUCDR~\cite{zhu2022personalized} provides another new perspective to formalize the cross-domain cold start problem as a meta-learning pipeline and build a personalized interest transfer bridge.

Both our work and PTUCDR employ parametric meta-learning pipelines for designing models, yet they exhibit significant differences.
Unlike PTUCDR, \textit{CVPM}'s meta-learner generates a bias network instead of a personalization bridge, and we leverage the bias network together with a common mapping function for personalized interest transfer.
Furthermore, we describe more fine-grained user representations in a probabilistic form as well as exploit the knowledge of non-overlapping users, both not available in PTUCDR. These differential measures overcome the pre-existing weaknesses mentioned in Section~\ref{sec:introduction}.

\subsection{Meta Learning}
Meta-learning, or learning to learn, is a technique that enables rapid acquisition of new skills based on past experience and a small number of samples~\cite{hospedales2021meta, wang2021preference}.
The prevailing meta-learning approaches fall into three genres, namely \textit{metric-based, model-based, optimization-based}~\cite{huisman2021survey}. 
Our work falls into the category of optimization-based methods, where the goal is to produce meta-parameters that are integrated with user representations to create personalized target representations.
MAML~\cite{finn2017model} is a classic example of this approach, as it involves constructing multiple sets of tasks to iteratively optimize meta-learners and base learners, thereby incorporating prior knowledge into initialization parameters.
Reptile~\cite{nichol2018reptile} simplifies the bi-optimization method into a first-order process, and strives to perform gradient descent by sampling multiple times for each task.
LEO~\cite{rusu2018meta} further introduces the representation network into the meta-learning framework, which alleviates the over-fitting drawback of MAML in high-dimensional spaces.
There have been numerous studies that have extended these ideas to single-domain recommender systems~\cite{wang2022deep}. For instance, MELU~\cite{lee2019melu} separates the recommendation module into an embedding layer and a decision layer, where the update of the decision layer follows the optimization process of meta-learning. FLIP~\cite{liu2021intent} regards the session behavior of each user as an independent task, and divides the session into support set and query set to formalize the MAML training process, thereby enhancing the capture of short-term user preferences. CHAML~\cite{chen2021curriculum} extends meta learning process to POI recommendation scenarios, and improves recommendation performance through low-accuracy hard sample sampling for both city-level and user-level.

In a nutshell, most existing recommender systems consider how to formalize single-domain recommendation tasks into a meta-learning training paradigm, but it is challenging to directly apply these approaches to cross-domain scenarios due to domain gaps.
In contrast, we formalize user uniqueness in interest transfer as learning a bias network with meta-learning, which subtly utilizes multiple cross-domain mapping tasks to improve the personalization and performance of cross-domain recommendation.

\subsection{Self-supervised Learning}
The essential idea of self-supervised learning is to explore its own supervision signal from large-scale unsupervised data, and train the network through this structured supervision information, thereby learning valuable representations for downstream tasks~\cite{liu2022graph}. Existing research paradigms can be broadly categorized as \textit{generative} and \textit{contrastive}, with the former reconstructing the original input by designing a proxy task and the latter distinguishing the instances themselves in a high-order feature space~\cite{liu2021self}.
Our research belongs to the contrastive self-supervision genre, which can be further subdivided into \textit{negative example-based, clustering-based} and \textit{redundancy loss-based}. For instance, SimCLR~\cite{chen2020simple} and Moco~\cite{he2020momentum} construct samples into two views, and input different encoders for comparative learning, in which the views of the same sample are positive pairs, and vice versa. Swav~\cite{caron2020unsupervised} assumes that views originating from the same instance are consistent across multiple semantic category distributions, thus avoiding the high computational cost of negative sampling. Barlow Twins~\cite{zbontar2021barlow} constructs a cross-correlation matrix from the high-order representations of two branch views, encouraging the diagonal elements to be 1 and the others to be 0 to enhance the discrimination of the same instance.
Inspired by them, a host of recommendation algorithms~\cite{jing2023contrastive,yao2022contrastive} leverage contrastive views such as user-user~\cite{chen2022intent}, user-centroid~\cite{lin2022improving}, and random mask operations~\cite{liu2022graph} to enhance the semantics of entity representations.

We also creatively introduce two contrasting views, group-level and individual-level, in this study, to augment the discriminative ability and mine the knowledge of non-overlapping users. This approach not only naturally extends the single-domain contrastive learning strategy to cross-domain scenarios, but also provides a new contrastive perspective under cross-domain mapping.

\section{PRELIMINARIES}\label{section3}
In this section, we first introduce the mathematical annotations throughout the paper and the definition of cross-domain recommendation. Then, we summarize the traditional training and testing pipelines of the \textit{embedding and mapping} genre as preliminaries.
\subsection{Notations and Problem Definition}
Suppose we have two domains, both of which contain user rating data, called source domain $\mathcal{D_S}$ (rich) and target domain $\mathcal{D_T}$ (sparse) according to the intensity of interaction. 
For two domains we have $\mathcal{D_{S}} = \{\mathcal{U_{S}, V_{S}, R_{S}}\}$, $\mathcal{D_{T}} = \{\mathcal{U_{T}, V_{T}, R_{T}}\}$ , where $\mathcal{U, V, R}$ refer to the user set, item set and observed interaction, respectively. 

\noindent\textbf{Cross-domain recommendation (CDR):} CDR overlaps on the user side, where users with ratings in both domains are called overlapping users $\mathcal{U}_{o}$, while users who are only active in a single domain are called non-overlapping users, i.e. $ \mathcal{U}_{s }$ and $\mathcal{U}_{t}$. Obviously $\mathcal{U_{S}}=\mathcal{U}_{s} \cup \mathcal{U}_{o}$, $\mathcal{U_{T}} = \mathcal{U}_{t} \cup \mathcal{U}_{o}$. CDR strives to learn the user mapping function $\mathcal{F}_{u}(\cdot)$ from the source domain to the target domain, thereby providing high-quality recommendations in the target domain.

\noindent\textbf{Cross-system recommendation (CSR):} CSR overlaps on the item side, i.e, $\mathcal{V_{S}}=\mathcal{V}_{s} \cup \mathcal{V}_{o}$, $\mathcal{V_{T}}=\mathcal{V}_{t} \cup \mathcal{V}_{o}$, where $\mathcal{V}_{s}, \mathcal{V}_{t}$ are non-overlapping items in the source domain and target domain respectively, and $\mathcal{V}_{o}$ is the overlapping thing. The goal of CSR is to find $\mathcal{F}_{v}(\cdot)$ for item representation mapping, but it is different from CDR in that it uses items as a bridge for knowledge transfer. Technically, during training, CSR only needs to swap the roles of users and items in the transfer strategy, and the other pipelines are consistent.

For convenience, unless otherwise stated, we will use the CDR as the main line for mathematical formalization.

\subsection{Traditional Embedding and Mapping Approach}\label{tra}
To facilitate readers to understand the background and clarify the technical contribution of the proposed algorithm, we summarize the traditional embedding and mapping pipeline as preliminary.
The general practice usually consists of three stages: \textit{embedding pre-training, mapping function learning}, and \textit{recommendation module}.
\subsubsection{Stage A: Embedding learning stage}\label{stageA}
In this stage, they pre-train latent factor models in source and target domains respectively to obtain semantic embeddings of users and items. Commonly used latent factor models include matrix decomposition (MF)~\cite{chen2018matrix}, singular value decomposition (SVD)~\cite{yuan2019singular}, generalized matrix decomposition (GMF)~\cite{he2017neural}, etc. Without loss of generality, MF is used here for pre-training. Formally, for each domain,
\begin{equation}
\hat{r}_{i j} = f(\mathbf{u}_{i}, \mathbf{v}_{j}),
\end{equation}
where $\mathbf{u}_{i}\in \mathbb{R}^{d}$ and $\mathbf{v}_{j}\in \mathbb{R}^{d}$ refer to the embeddings of user $u_{i}$ and item $v_{j}$, and $\hat{r}$ is estimated score.
$f(\cdot)$ can be an inner product, or a deep learning module, such as a multi-layer perceptron (MLP).
\begin{equation}
\mathcal{L}(\hat{r}_{i j}, r_{i j})=\frac{1}{|\mathcal{D}|} \sum_{r_{i j} \in \mathcal{D}}(\hat{r}_{i j}-r_{i j})^2 + \lambda(\sum_{i}||\mathbf{u}_{i}|| + \sum_{j}||\mathbf{v}_{j}||),
\end{equation}
where $\lambda$ is the coefficient of the regularization term, and $||\cdot||$ represents the 2-norm.
In this way, they learns the latent matrices $\mathbf{U}_{\mathcal{S}} \in \mathbb{R}^{|\mathcal{U}_{\mathcal{S}}| \times d}$, $\mathbf{V}_{\mathcal{S}} \in \mathbb{R}^{|\mathcal{V}_{\mathcal{S}}| \times d}$ from $\mathcal{R_{S}}$ and $\mathbf{U}_{\mathcal{T}}\in \mathbb{R}^{|\mathcal{U}_{\mathcal{T}}| \times d}$, $\mathbf{V}_{\mathcal{T}} \in \mathbb{R}^{|\mathcal{V}_{\mathcal{T}}| \times d}$ from $\mathcal{R_{T}}$.

\subsubsection{Stage B: Mapping function learning stage}\label{stageB}
In the second stage, once the semantic representations of the two domains are obtained, overlapping users are exploited to train the mapping function. The mapping function here can be linear or non-linear. Formally,
\begin{equation}
\label{equ1}
\mathcal{L}=\sum_{u_i \in \mathcal{U}_o}\|g(\mathbf{u}_i^s ; \theta)-\mathbf{u}_i^t\|,
\end{equation}
where $g(\cdot)$ denotes a common mapping function, and $\theta$ is its corresponding parameter.

\subsubsection{Stage C: Recommendation stage}\label{stageC}
In the third stage, when the loss tends to converge, $g(\cdot)$ can be used for representation transfer between domains.
\begin{equation}
    \mathbf{\hat{u}}_{i}^{t} = g(\mathbf{u}_{i}^{s};\theta),
\end{equation}
where $\mathbf{\hat{u}}_{i}^{t}$ denotes the transferred target domain representation of $u_{i}^{s}$, and it can be applied to calculate the matching score with $\mathcal{V_T}$ for recommendation.

Through these three stages, the model acquires a common mapping function by leveraging overlapping users, enabling cross-domain interest transfer and facilitating recommendations for cold-start users in the target domain.
Despite the decent performance, the coarse-grained user characterization, impersonal mapping function, and over-reliance on overlapping users create barriers for its widespread application.
These three weaknesses present opportunities for us to further enhance the genre, and the targeted improvement measures we propose in this paper are the primary technical contributions.

\section{Methodology}\label{section4}
\begin{figure*}[!t]
	\centering
\includegraphics[width=\linewidth,height=0.4\linewidth]{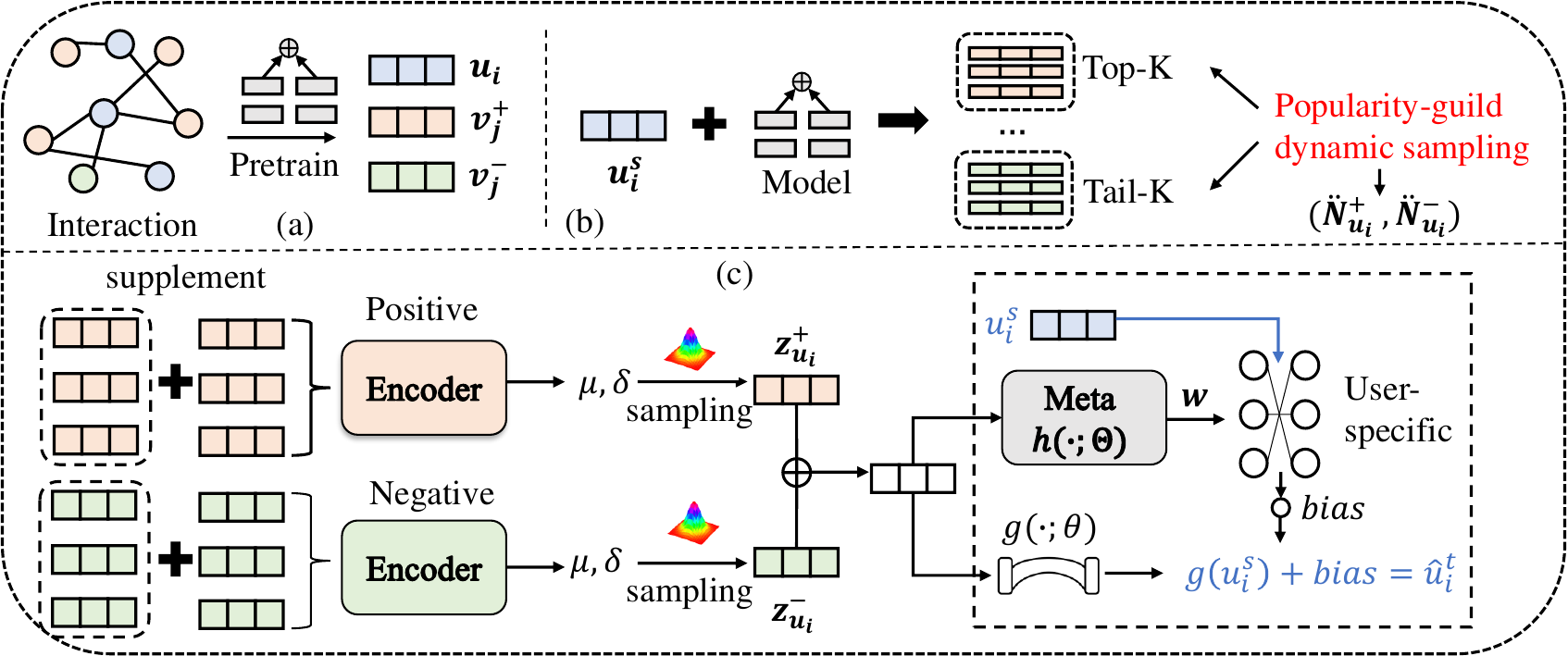}
	\caption{The overall framework of \textit{CVPM}. (a) Pretraining using MF. (b) Sampling enhancment. (c) Distribution learning. (d) Generic bridge and personalized bias network. Blue, orange, and green represent users, positive items, and negative items, respectively.}
	\label{frame}
\end{figure*}

In this section, we elaborate on each sub-module of \textit{CVPM} and clarify the significance of its design. Fig.~\ref{frame} provides a clear and intuitive overview of the algorithm flow.

\subsection{Valence Preference Learning}
Similar to \textit{Stage A}, we first need to obtain transferable user representations.
However, considering that the latent semantic representation generated in \textit{Stage A} cannot distinguish fine-grained user preferences, we refer to the valence preference theory~\cite{liu2022review,kato2022rating, kumar2020movie,ullah2016valence} to represent user interests from both positive and negative perspectives. 
Specifically, we obtain the pre-trained $\mathbf{U}_{\mathcal{S}},\mathbf{V}_{\mathcal{S}},\mathbf{U}_{\mathcal{T}},\mathbf{V}_{\mathcal{T}}$
 as in \textit{Stage A}, and then divide each user's interacted items in the source domain into positive and negative categories for preference characterization. A naive approach is to divide according to the median rating of each user, and then perform average pooling for aggregation.
 Formally,
\begin{equation}
\begin{split}
    \mathbf{p}^{+}_{u_i} &= \frac{1}{|\mathcal{N}_{u_{i}}^{+}|} \sum_{v_{j} \in \mathcal{N}_{u_{i}}^{+}, r_{ij} \geq M_{u_i}} \mathbf{v}_j \\
    \mathbf{p}^{-}_{u_i} &= \frac{1}{|\mathcal{N}_{u_{i}}^{-}|} \sum_{v_{j} \in \mathcal{N}_{u_{i}}^{-}, r_{ij} < M_{u_i}} \mathbf{v}_j
\end{split},
\end{equation}
where $M_{u_{i}}$ refers to the median of $u_{i}$ scores, $\mathcal{N}^{+}$ and $\mathcal{N}^{-}$ refer to the set of items with ratings above and below the median, respectively. $|\cdot|$ denotes set size, $\mathbf{p}^{+}$ and $\mathbf{p}^{-}$ represent the user's positive preference and negative preference respectively.
Despite the simplicity and effectiveness, pooling aggregation may lead to information loss and sensitivity to noisy data~\cite{zhang2019depth}. Consequently, we design a learning paradigm of probability distributions to represent valence preferences, including two steps of heuristic sampling and distribution learning.

\noindent\textbf{Sampling enhancement:}\label{sampling}
Distributional learning with $\mathcal{N}^{+}$ and $\mathcal{N}^{-}$ is great if user interaction is rich at both levels. Nevertheless, in practical tasks, many users' interactions are relatively sparse, underrepresenting their positive and negative preferences. On account of this, we utilize the pretrained user embeddings to sample several pseudo-interaction items as a complement to positive and negative distribution learning.
First, we use the trained embedding to estimate the $u_{i}$ ratings on all items $\mathcal{V_{S}} \setminus (\mathcal{N}^{+}_{u_{i}}  \cap \mathcal{N}^{-}_{u_{i}})$ and sort them in descending order, and then intercept the first $|\mathcal{\dot{N}}^{+}_{u_{i}}|$ and last $|\mathcal{\dot{N}}^{-}_{u_{i}}|$ items as the user's positive and negative preference candidates. 
Then, we adopt different heuristic sampling strategies for $\dot{\mathcal{N}}^{-}_{u_{i}}$ and $\dot{\mathcal{N}}^{+}_{u_{i}}$ to complement positive and negative preferences. Considering that popular items are easily known by users, it is reasonable to think that missed popular items are more likely to belong to negative preferences~\cite{chen2023revisiting}, so we use the reciprocal of candidate popularity as the probability to sample $\dot{\mathcal{N}}^{-}_{u_{i}}$.
In contrast, driven by popularity, users may have consistent behavior when expressing positive preferences, therefore we use tf-idf values~\cite{aizawa2003information} as probabilities to sample $\dot{\mathcal{N}}^{+}_{u_{i}}$ , to introduce more personalized items. Formally,

\begin{equation}
p^{i,-}_{v_{j}} = Norm(1/\frac{n_{v_j}}{\sum_{j=1}^{|\dot{\mathcal{N}}^{-}|}n_{v_j}}),
\end{equation}
\begin{equation}
p^{i,+}_{v_{j}} = Norm(\frac{n_{v_j}}{\sum_{j=1}^{|\dot{\mathcal{N}}^{+}|}n_{v_j}} * lg\frac{|\dot{\mathcal{N}}^{+}|}{|\{j: v_{j} \in \dot{\mathcal{N}}^{+}_{u_{i}} \}|}),
\end{equation}
where \textit{Norm} refers to $\frac{1}{1+\exp (-x)}$, $n_{v_{j}}$ refers to the number of $v_{j}$ appearing in the data set, and $p^{i}_{v_{j}}$ represents the sampling probability of user $u_i$ to $v_j$.

\begin{figure}[!h]
	\centering
	\includegraphics[width=\linewidth,height=0.4\linewidth]{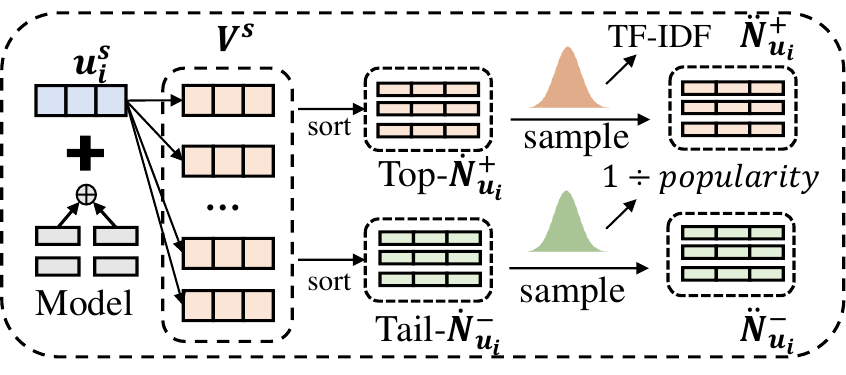}
	\caption{Popularity-guild dynamic sampling. We adopt different heuristic sampling strategies for positive and negative candidate sets.}
	\label{smaple}
\end{figure}

According to the obtained user-item probability matrix, we can sample $|\mathcal{\ddot{N}}_{u_{i}}^{+}|$ positive items, $|\mathcal{\ddot{ N}}_{u_{i}}^{-}|$ negative items to complement the user interaction set before each iteration, i.e. $\mathcal{N}_{u_{i}}^{+} = \mathcal{N}_{u_{i}}^{+} \cup \mathcal{\ddot{N}}_{u_{i}}^{+}$, $\mathcal{N}_{u_{i}}^{-} = \mathcal{N}_{u_{i}}^{-} \cup \mathcal{\ddot{N}}_{u_{i}}^{-}$. A concise framework can be found in Fig.~\ref{smaple}.

\noindent\textbf{Distribution learning:}
Unlike previous work that directly characterizes user representations, we aim to learn positive and negative preference distributions at a finer-grained level. 
First, we use the attention mechanism to obtain the representations of positive and negative preferences. Formally,
\begin{equation}
\mathbf{p}_{u_i}^{+}=\sum_{v_j \in \mathcal{N}_{u_{i}}^{+}} \alpha_j \mathbf{v}_{j}, \qquad
\mathbf{p}_{u_i}^{-}=\sum_{v_j \in \mathcal{N}_{u_{i}}^{-}} \beta_j \mathbf{v}_{j},
\end{equation}
where $\alpha_{j}=\frac{exp(\mathbf{v}_{j}W_{1})}{\sum_{v_{j}\in\mathcal{N}_{u_{i}}^{+}}exp(\mathbf{v}_{j}W_{1})}$, $\beta_{j}=\frac{exp(\mathbf{v}_{j}W_{2})}{\sum_{v_{j}\in\mathcal{N}_{u_{i}}^{+}}exp(\mathbf{v}_{j}W_{2})}$, and $W$ denotes trainable parameters.
Then, without loss of generality, we assume that the two preferences both follow Gaussian distribution and adopt reparameterization trick to express the random variable, i.e., $z_i \sim {N}(\mu_{i}, \operatorname{diag}(\sigma_i^2))$. Formally, we can obtain the positive distribution $\mathbf{z}_{u_{i}}^{+}$ as follows,
\begin{equation}
\label{equ2}
\begin{gathered}
\mathbf{p}_{u_{i}}^{+}=\operatorname{ReLU}({W}_{+}^{1} \mathbf{p}_{u_{i}}^{+}), \\
\mu_i=W_{\mu}^{1}  \mathbf{p}_{u_{i}}^{+}, \log {\sigma}_i=W_{\sigma}^{1} \mathbf{p}_{u_{i}}^{+}, \\
\mathbf{z}_{u_{i}}^{+}={\mu}_i+{\epsilon} \odot {\sigma}_i, {\epsilon} \sim {N}(0, \mathcal{I}),
\end{gathered}
\end{equation}
where $\epsilon$ is the Gaussian noise.
Likewise, we can obtain the negative preference $\mathbf{z}_{u_{i}}^{-}$ in the same way,
\begin{equation}
\label{equ3}
\begin{gathered}
\mathbf{p}_{u_{i}}^{-}=\operatorname{ReLU}({W}_{-}^{2} \mathbf{p}_{u_{i}}^{-}), \\
\mu_i=W_{\mu}^{2} \mathbf{p}_{u_{i}}^{-}, \log {\sigma}_i=W_{\sigma}^{2} \mathbf{p}_{u_{i}}^{-}, \\
\mathbf{z}_{u_{i}}^{-}={\mu}_i+{\epsilon} \odot {\sigma}_i, {\epsilon} \sim {N}(0, \mathcal{I}).
\end{gathered}
\end{equation}
To sum up, we discriminatively learn users' positive and negative preferences via probabilistic representations. It not only improves the integrity and diversity of transferable representations but also bolsters their resilience against noise interference.

\subsection{Common Bridge and Personalized Biases}
Once transferable representations are obtained, \textit{Stage B} puts efforts on training a common mapping function that is shared among all users.
We contend that this practice violates the personalized nature of recommender systems, which fail to observe subtle differences among users, resulting in an approximation of all user representations.
Given this, we resort to meta-learning to learn a personalized bias parameter for each user, which together with the common mapping function constitutes the user's representation in the target domain.

\noindent\textbf{Mapping function:}
This module learns the traditional mapping function $g(\cdot)$. We first fuse users' positive and negative preferences in the source domain into user representations, and then perform mapping.
\begin{equation}
    \mathbf{p}^{s}_{u_{i}} = \mathbf{z}_{u_{i}}^{+} \oplus \mathbf{z}_{u_{i}}^{-},
\end{equation}
\begin{equation}
    \mathbf{\tilde{u}}_{i}^{t} = g_(\mathbf{p}^{s}_{u_{i}};\theta),
\end{equation}
where $\oplus$ denotes concatenation operation. This transformation captures the common user transformation schema.

\noindent\textbf{User-specific bias:}
In addition to the common mapping function, we also build a meta-learner $h$ to learn a bias network to preserve user personalization. Formally,
\begin{equation}
    \mathbf{w} = h(\mathbf{p}^{s}_{u_i}; \Theta),
\end{equation}
where $\Theta$ refers to the parameters of the meta-learner $h$. Please note that instead of outputting a scalar bias, a network parameter $\textbf{w}$ is output here, which has the ability to learn the user's personalized \textit{bias}. 
Next, we input $u_{i}^{s}$ to generate personalized transfer bias.
\begin{equation}
    bias_{u_{i}} = \mathbf{w} * \mathbf{u}_{i}^{s},
\end{equation}
where $bias_{u_{i}}$ is a specific parameter for $u_i$.
\begin{equation}
    \hat{u}_{t} = \mathbf{\tilde{u}}_{i}^{s} + bias_{u_{i}}.
\end{equation}
In generall, compared with the previous $\mathcal{F}_{u}(\cdot)=g(\cdot;\theta)$, we expand it to $\mathcal{F}_{u}(\cdot)=g (\cdot;\theta)+h(\cdot; \Theta)$. 
In this manner, we are capable of perceiving global transfer patterns from common mapping functions as well as taking into account users' personalized habits.

\subsection{Model Optimization}
\textbf{Overlapping user exploitation: }Previous studies have focused on narrowing the distance between the source and target domains using Eq.~\ref{equ1}, but this approach may not be effective for users with less interaction. Specifically, users with less interaction cannot obtain accurate user representation during the pre-training stage, let alone accurate transfer. Therefore, we train using user interactions in the target domain as supervision targets to ensure safe transfer.
\begin{equation}
\label{equ4}
\mathcal{L}_{o} = \frac{1}{|\mathcal{R}_o^t|} \sum_{r_{i j} \in \mathcal{R}_o^t}(r_{i j}-\mathcal{F}_{u_i}(\mathbf{u}_i^s ;\mathbf{p}_{u_i}^{s}, \theta, \Theta) \mathbf{v}_j)^{2},
\end{equation}
where $\mathcal{R}_o^t$ is the set of overlapping user ratings in the target domain. Meanwhile, this optimization way can also exert its ability in warm start scenario, which we will reflect in the Section~\ref{section6}.

\noindent\textbf{Non-overlapping user exploitation: }As discussed in section~\ref{sec:introduction}, using only information from overlapping users may bias the model and cause loss of knowledge, especially if their proportion is low. Consequently, we strive to mine the knowledge embedded in non-overlapping user interactions, which is rarely considered. In particular, we learn instance discriminants at the group and individual-levels to enhance the semantics and specificity of representations.

For the group level, we encourage users' transferred target domain representations to be closer to the semantic centroid of their affiliation in the source domain. This way can convey similarity relations and semantic preferences among users.

\begin{equation}
\mathcal{L}_{ng} =\sum_{u_{i} \in \mathcal{U}_S}-\log \frac{\exp (\hat{\mathbf{u}}_i^t \cdot \mathbf{c}_{u_{i}}^{s} / \tau)}{\sum_{c_{i} \in \mathcal{C}} \exp (\mathbf{u}_i^s \cdot \mathbf{c}_{i} / \tau)},
\end{equation}
where $\mathcal{C}$ refers to the set of centroids obtained using k-means, $\mathbf{c_{u_{i}}^{s}}$ refers to the centroid representation to which $u_i$ belongs, and $\tau$ refers to the temperature coefficient. 

For the individual level, we naturally match the positive and negative preferences obtained in Eq.~\ref{equ2} and Eq.~\ref{equ3} with user representations as positive and negative pairs. This approach helps the model distinguish between user-personalized positive and negative preferences, and facilitates the transfer of non-overlapping user knowledge. Formally,

\begin{equation}
\mathcal{L}_{ni} = max(||\mathbf{u}_i^{t}-\mathbf{z}_{u_i}^{+}||^{2} - ||\mathbf{u}_i^{t}-\mathbf{z}_{u_i}^{-}||^2 + margin, 0),
\end{equation}
where \textit{margin} is used to control the minimum distance between positive and negative pairs. Here we set it to 0.001 by tuning.

\begin{figure}[!ht]
	\centering
	\includegraphics[width=\linewidth,height=0.4\linewidth]{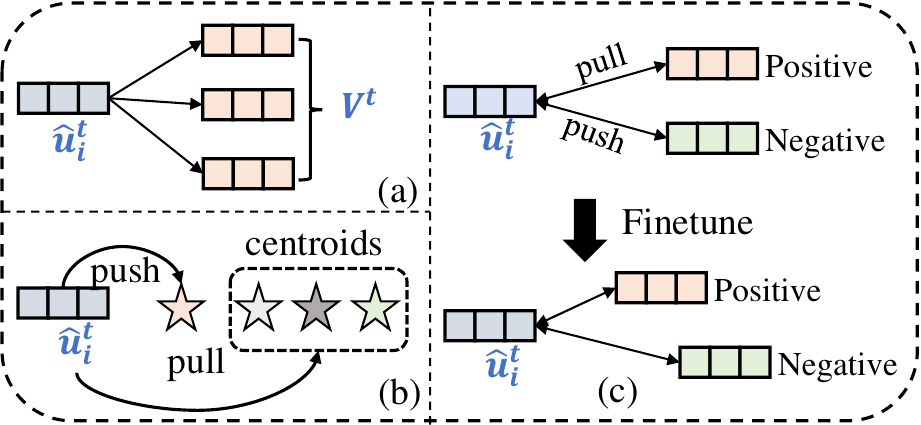}
	\caption{Model optimization. (a) Supervision signals for overlapping users. (b) Group-level contrastive loss. (c) Individual-level contrastive loss. Stars of different colors represent the cluster centroids of the source domain.}
	\label{opt}
\end{figure}

A concise framework for the aforementioned three losses can be seen in Fig.~\ref{opt}. We optimize the aforementioned three losses through a weighted sum approach, formally,
\begin{equation}
\mathcal{L}_{total} = \mathcal{L}_{o} + \gamma(\mathcal{L}_{ng} + \mathcal{L}_{ni}),
\end{equation}
where $\gamma$ is the trade-off between supervised and self-supervised signals.

In short, this paper leverages meta-learning and self-supervised learning to address the aforementioned three weaknesses, thereby kicking the cross-domain recommendation framework of \textit{embedding and mapping} genre to higher stairs. See Algorithm~\ref{alg} for a clear algorithm flow.
\begin{algorithm} 
\begin{small}
\caption{The Algorithm of \textit{CVPM}} 
\label{alg} 
\begin{algorithmic}[1] 
\REQUIRE $\mathcal{D_{S}}$,  $\mathcal{D_{T}}$, maximum epoch $N$; 
\ENSURE Parameters $\theta$, $\Theta$;

\STATE \textbf{Pretraining Stage.} User and item embedding training for source and target domains.
$$\min \mathcal{L}(r_{i,j}, \hat{r}_{i,j}),$$
\FOR{epoch i $\leftarrow$ 1 to N:}
\STATE Sample a batch of u
\STATE \textbf{Embedding Learning Stage.} Sampling items to complement positive and negative sets,
$\mathcal{N}_{u_{i}}^{+} = \mathcal{N}_{u_{i}}^{+} \cup \mathcal{\ddot{N}}_{u_{i}}^{+}, \mathcal{N}_{u_{i}}^{-} = \mathcal{N}_{u_{i}}^{-} \cup \mathcal{\ddot{N}}_{u_{i}}^{-}$. and then perform probabilistic representation learning,
$$\mathbf{z}_{u_{i}}^{+}={\mu}_i+{\epsilon} \odot {\sigma}_i, \qquad
\mathbf{z}_{u_{i}}^{-}={\mu}_i+{\epsilon} \odot {\sigma}_i$$
\STATE \textbf{Mapping Learning Stage}. Train the mapping function and bias meta-network
$$\min _{\theta, \phi} \mathcal{L}_{o}+\gamma(\mathcal{L}_{ng}+\mathcal{L}_{ni}) $$
\ENDFOR
\STATE Update the parameters
\RETURN Parameters $\theta$, $\Theta$

\STATE \textbf{Recommendation Stage} For cold start scenario, use the trained mapping function $\mathcal{F}(\cdot; \theta, \Theta)$ as follows. 
$$\hat{u}^{t}_{i} = \mathcal{F}_{u_i}(\mathbf{u}_i^s ;\mathbf{p}_{u_i}^{s}, \theta, \Theta)$$
For warm start scenarios, fine-tuning with Eq.~\ref{equ4} is required before mapping.
\end{algorithmic}
\end{small}
\end{algorithm}

\section{Experiments and Analysis}\label{section5}
In this section, we first introduce the experimental settings, and then present and analyze the experimental results.
\subsection{Experimental Settings}
We elaborate on experimental details and compared baselines. For the sake of fairness, the experimental settings of the comparison methods are all in accordance with the original paper, unless otherwise specified.

\begin{table}
\centering
\caption{Statistics of data sets.}
\label{statis}
\begin{tabular}{c|c|cccc} 
\hline
\multicolumn{2}{c|}{Data set}         & \#Users & \#Items & \#Ratings & Sparsity  \\ 
\hline
\multirow{3}{*}{Amazon} & MoviesTV    & 123,960 & 50,052  & 1,697,533 & 0.9997    \\
                        & CDsVinyl    & 75,258  & 64,443  & 1,097,592 & 0.9998    \\
                        & Books       & 603,668 & 367,982 & 8,898,041 & 0.9999    \\ 
\hline
\multirow{3}{*}{Douban} & DoubanMovie & 2,712   & 34,893  & 1,278,401 & 0.9865    \\
                        & DoubanBook  & 2,212   & 95,872  & 227,251   & 0.9989    \\
                        & DoubanMusic & 1,820   & 79,878  & 17,9847   & 0.9988    \\ 
\hline
Movielens               & Subset      & 35,815  & 10,998  & 6,855,257 & 0.9826    \\
Netflix                 & Subset      & 37,769  & 3,816   & 2,170,403 & 0.9849    \\
\hline
\end{tabular}
\begin{tabular}{c|c|ccc|c} 
\toprule
\multicolumn{2}{c|}{Tasks}   & Source      & Target      & \#Overlap & S-Ratio \\ 
\hline
\multirow{5}{*}{CDR} & Task1 & MoviesTV    & CDsVinly    & 18,031    & 0.1454   \\
                     & Task2 & Books       & MoviesTV    & 37,388    & 0.0619   \\
                     & Task3 & Books       & CDsVinly    & 16,738    & 0.0277   \\ 
\cline{2-6}
                     & Task4 & DoubanMovie & DoubanBook  & 2,209     & 0.8145   \\
                     & Task5 & DoubanMovie & DoubanMusic & 1,815     & 0.4369   \\ 
\hline
CSR                  & Task6 & Movielens   & Netflix     & 685       & 0.0622   \\
\bottomrule
\end{tabular}
\end{table}

\begin{figure} \centering   
\subfigure[Overlapping ratio]{ 
\label{ana:or}     
\includegraphics[width=0.28\linewidth,height=0.35\linewidth]{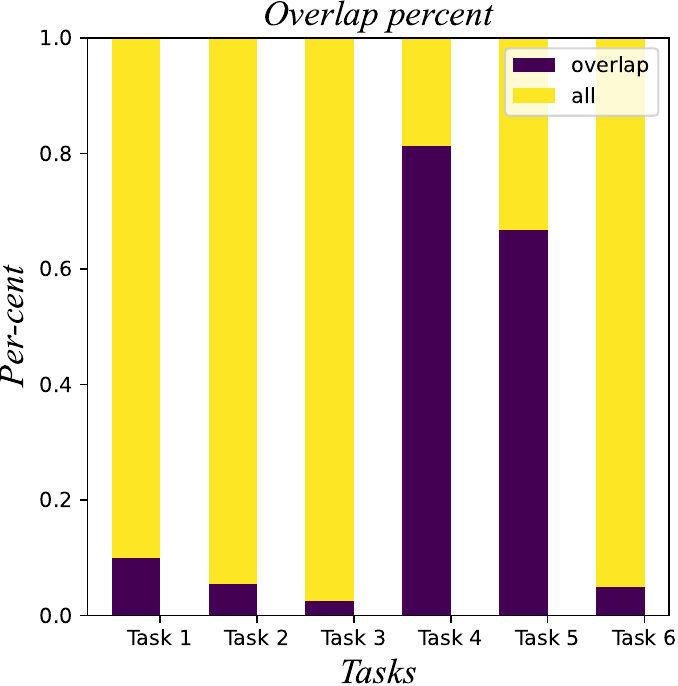}
}    
\subfigure[Valence distribution] { 
\label{ana:vd}     
\includegraphics[width=0.28\linewidth,height=0.35\linewidth]{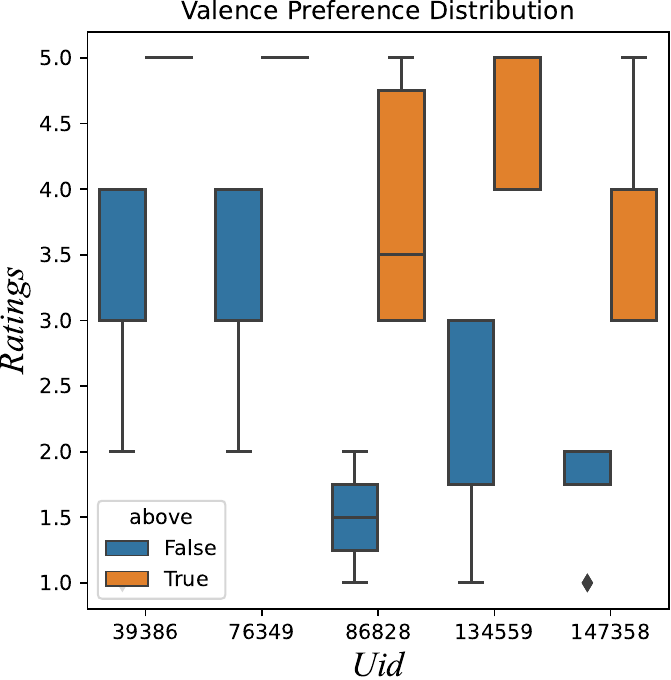}
}   
\subfigure[Ratings distribution] { 
\label{ana:vr}     
\includegraphics[width=0.28\linewidth,height=0.35\linewidth]{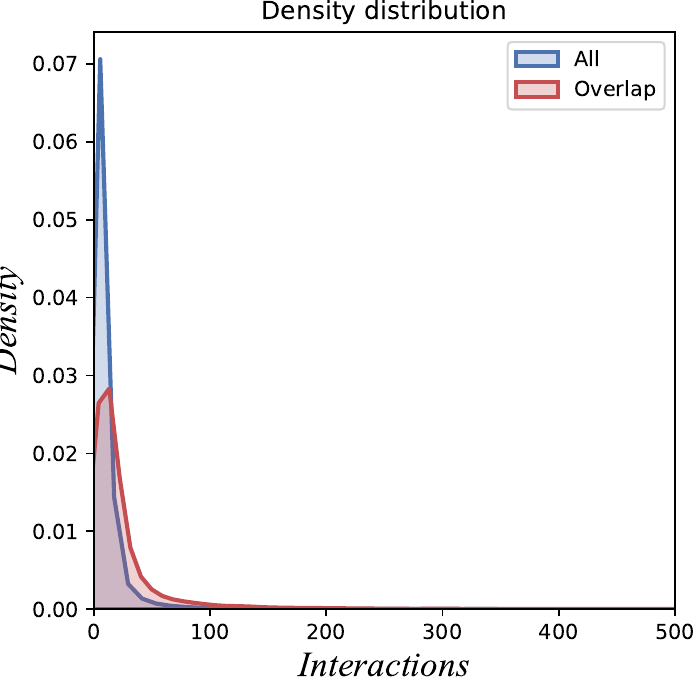}
}    
\centering
\caption{Data analysis.}
\label{ana}     
\end{figure}

\subsubsection{Data sets \& analysis}
We construct 5 cross-domain recommendation tasks and 1 cross-system recommendation task based on 8 data sets to evaluate the performance of \textit{CVPM} and all compared baselines. Table~\ref{statis} presents statistics for data sets and built tasks. Among them, MoviesTV, CdsViny, and Books come from three subcategories of Amazon plateform~\footnote{https://www.amazon.con/}, each containing a wealth of user rating data.
Likewise, DoubanMoive, DoubanBook and DoubanMusic are crawled from Douban~\footnote{https://www.douban.com/} website. We also sampled a subset of each from Movielens~\footnote{https://grouplens.org/data sets/movielens/} and Netflix~\footnote{https://www.netflix.com/} public data sets for experiments. The user IDs of these data sets from the same website are common, and we exploit this pairwise combination to build cross-domain recommendation tasks, i.e., tasks 1-5. In particular, for the Movie-lens and Netflix data sets from different systems, we recode item IDs by movie names and build a cross-system recommendation task, i.e., task 6. Distinguishing between source and target domains depends on the number of interactions, and we assume that more frequent interactions imply more preference patterns and knowledge~\cite{zang2022survey}. S-ratio is the proportion of overlapping users in the source domain users, representing the proportion of transferable knowledge~\cite{kang2019semi}. It is evident that in the vast majority of tasks, the transferable knowledge accounts for less than 15

We also perform several preliminary data analysis to prove our intuition. 
Fig.~\ref{ana:or} shows the proportion of overlapping users to total users. Except for the tasks derived from Douban, the proportion of overlapping users is less than 10\% in cross-domain tasks.
This demonstrates that overlapping users alone are insufficient to cover all transfer patterns, highlighting the importance of leveraging knowledge from non-overlapping users.
Fig.~\ref{ana:vd} presents a boxplot of positive and negative ratings for 5 random user, with significantly different variances. This example exemplifies the contrast between the distribution of positive and negative user preferences. Fig.~\ref{ana:vr} presents the density distribution of user ratings, both overlapping users and overall users, showing a long tail. This proves the necessity of our sampling supplement in section~\ref{sampling}.


\subsubsection{Parameters settings}
Experiments were performed on an Ubuntu 18.04 server with an Intel(R) Xeon(R) Gold 5118 CPU (12 cores, 2.30GHz) and a single NVIDIA Tesla V100 GPU. PyTorch was used as the framework for efficient model training and inference, harnessing the power of the GPU's parallel computing capabilities. We set batch sizes of 512 and 256 in the pre-training and meta-learning phases, respectively, and use the Adam optimizer with a learning rate of 1e-2 and a decay of 5e-4 for gradient descent. The embedding size $d$ we use is $[5,10,20,30]$, the number of centers $|\mathcal{C}|$ is $[50,100,150,200]$, and the number of supplementary samples $|\ddot{\mathcal{N}}|$ is $[5,10,15,20]$, and the auxiliary task weight $\gamma$ is $[1e-2, 5e-3, 1e-3, 1e-4]$, see section~\ref{tune} for tuning.

\noindent \textbf{Baselines methods:}
Similar to~\cite{zhu2021transfer,kang2019semi,man2017cross}, we mainly select several state-of-the-art algorithms based on the \textit{embedding and mapping genre} for comparison to demonstrate our advancement and effectiveness.

\begin{itemize}[leftmargin=12pt]
    \item \textbf{TGT~\cite{chen2018matrix}:} TGT performs MF decomposition directly in the target domain without cross-domain design.
    \item \textbf{CMF~\cite{singh2008relational}:} CMF shares user embeddings in both domains for knowledge transfer.
    \item \textbf{CST~\cite{pan2010transfer}:} CST uses the embedding trained in the source domain to initialize the users and items shared by the target domain and the source domain, and then uses the data of the target domain for fine-tuning.
    \item \textbf{EMCDR~\cite{man2017cross}:} EMCDR puts effort into exploiting overlapping users to learn a common mapping function for transferring source-to-target representations.
    \item \textbf{DCDCSR~\cite{zhu2020deep}:} DCDCSR leverages the frequency of user interaction in the two domains to construct offset coefficients, thereby building a more robust cross-domain recommender system.
    \item \textbf{SSCDR~\cite{kang2019semi}:} SSCDR strives to preserve triangular correlations during transfer by constructing a semi-supervised loss with non-overlapping user knowledge.
    \item \textbf{TMCDR~\cite{zhu2021transfer}:} TMCDR introduces the idea of MAML into the cross-domain recommendation system, which learns how to transfer interests from the behavior of overlapping users by sampling multiple tasks.
    \item \textbf{LACDR~\cite{wang2021low}:} LACDR employs autoencoder to perform dimensionality reduction before mapping to alleviate the over-fitting phenomenon caused by embedding size.
    \item \textbf{PTUCDR~\cite{zhu2022personalized}:} PTUCDR utilizes the user's behavior in the source domain to construct a personalized bridge for each user.
\end{itemize}
For fairness, all baselines are carefully fine-tuned to ensure the best performance.

\noindent\textbf{Evaluation metrics:} Following~\cite{man2017cross,wang2021low,kang2019semi}, we randomly select a part of overlapping users, remove all his interactions and treat it as a test set. In our experiments, the train-test ratios of overlapping users are 80\%:20\%, 50\%:50\% and 20\%:80\%, respectively. For the warm-start scenario, half of the interactions of each test user are reserved for fine-tuning and the other half for testing. We divide strictly by time to avoid feature crossing. To evaluate the performance of our proposed model, we employ widely used metrics in cross-domain recommendation, i.e., Mean Absolute Error (MAE), Root Mean Square Error (RMSE), Hit Ratio (Hit) and Normalized Discounted Cumulative Gain (NDCG). MAE and RMSE measure the value estimation ability, and Hit and NDCG measures the top-k ranking ability of the model~\footnote{Note that for top-k evaluation, to prevent over-fitting, we sample 4 negative samples for each interaction. Negative samples are randomly selected from items that the user has never interacted with.}. For MAE and RMSE, the smaller the value, the better the performance; for Hit and NDCG, the larger the value, the better the performance. For MAE and RMSE, smaller values indicate better performance, while for Hit and NDCG, larger values indicate better performance, denoted by $\downarrow$ and $\uparrow$, respectively.

\subsection{Experiments Results}
In this subsection, we examine the performance of all baselines and the proposed \textit{CVPM} on 5 cross-domain recommendation tasks and 1 cross-system recommendation task, respectively. In particular, we also conduct experiments for different train-test ratios.

\noindent\textbf{Comparison with other baselines:}
The evaluation results of all baselines and \textit{CVPM} on the 6 tasks are shown in Table~\ref{comp:value} and Table~\ref{comp:top-k}. Evidence shows that \textit{CVPM} outperforms the best baselines on all metrics across all tasks, even up to 15\%.
TGT is a single-domain baseline that only relies on the collaborative filtering signal of the target domain, and thus performs least well.
CMF and CST are dedicated to fusing source domain information at the data set or initialization level, which inevitably brings noise. Compared with the above three baselines, EMCDR has a great improvement, but its performance is limited by the design of the common mapping function, which cannot distinguish the individual differences of users well.
DCDCSR and LACDR are variants of EMCDR that enhance representation semantics by adding representation offset and reconstruction loss to the original training pipeline, thus improving up to 6\%-10\% in task 4-5. Nevertheless, their effectiveness in other tasks is significantly reduced as they fail to grasp sufficient user habits and transmission patterns when the absolute number of overlapping users decreases. SSCDR employs the transfer of user-item triangles to leverage non-overlapping user knowledge, but its subpar performance on task 4 highlights its susceptibility to noisy data. TMCDR and PTUCDR are meta-learning based baselines with performance close to \textit{CVPM} on tasks 2-4. Nonetheless, the employment of a coarse-grained user representation coupled with the disregard for non-overlapping users significantly restricts their efficacy in task 1 and 6, resulting in 6\% and 10\% lower value estimation and top-k ranking, respectively, compared to \textit{CVPM}.

In a nutshell, \textit{CVPM} achieves the strongest performance on all tasks, both in terms of valuation and ranking ability. Our approach effectively surmounts the limitations of existing baselines, propelling the cross-domain recommendation of the embedded mapping genre to new heights.

\begin{table*}
\centering
\caption{Performance comparison: value estimation}
\label{comp:value}
\begin{tabular}{ccccccccccccc} 
\toprule
Task                    & Train-ratio           & Metric            & TGT    & CMF    & CST    & EMCDR  & DCDCSR & SSCDR  & TMCDR  & LACDR  & PTUCDR & CVPM             \\ 
\hline
\multirow{6}{*}{Task-1} & \multirow{2}{*}{80\%} & MAE $\downarrow$  & 4.5019 & 1.4722 & 1.8594 & 1.3596 & 1.7739 & 1.0910 & 1.0888 & 1.5161 & 1.1122 & \textbf{1.0272}  \\
                        &                       & RMSE $\downarrow$ & 5.1879 & 1.9636 & 2.2441 & 1.6445 & 2.0237 & 1.3894 & 1.4867 & 1.9681 & 1.4567 & \textbf{1.3280}  \\
                        & \multirow{2}{*}{50\%} & MAE $\downarrow$  & 4.4872 & 1.6221 & 2.0629 & 1.5760 & 1.9010 & 1.3335 & 1.4378 & 1.6784 & 1.2564 & \textbf{1.2339}  \\
                        &                       & RMSE $\downarrow$ & 5.2109 & 2.1762 & 2.4845 & 1.9087 & 2.1970 & 1.6810 & 1.8806 & 2.1475 & 1.6786 & \textbf{1.4936}  \\
                        & \multirow{2}{*}{20\%} & MAE $\downarrow$  & 4.5782 & 2.4855 & 2.1174 & 2.0959 & 2.1451 & 1.7770 & 1.9873 & 2.0419 & 1.5436 & \textbf{1.3429}  \\
                        &                       & RMSE $\downarrow$ & 5.3407 & 3.1382 & 2.6028 & 2.4474 & 2.4170 & 2.2347 & 2.3595 & 2.7405 & 2.1253 & \textbf{1.8169}  \\ 
\hline\hline
\multirow{6}{*}{Task-2} & \multirow{2}{*}{80\%} & MAE $\downarrow$  & 4.1955 & 1.4471 & 2.0896 & 1.1289 & 1.5474 & 1.0158 & 1.1332 & 1.2872 & 1.0541 & \textbf{0.9978}  \\
                        &                       & RMSE $\downarrow$ & 4.7992 & 1.8889 & 2.4840 & 1.4077 & 1.8034 & 1.2875 & 1.4783 & 1.5974 & 1.3695 & \textbf{1.2778}  \\
                        & \multirow{2}{*}{50\%} & MAE $\downarrow$  & 4.2853 & 1.6099 & 2.1461 & 1.2069 & 1.7179 & 1.0832 & 1.2394 & 1.3558 & 1.0956 & \textbf{1.0774}  \\
                        &                       & RMSE $\downarrow$ & 4.9206 & 2.0921 & 2.4601 & 1.5132 & 1.9776 & 1.4275 & 1.4593 & 1.6953 & 1.4413 & \textbf{1.4159}  \\
                        & \multirow{2}{*}{20\%} & MAE $\downarrow$  & 4.3221 & 1.7078 & 2.4930 & 1.3213 & 1.6889 & 1.2477 & 1.3984 & 1.8474 & 1.1874 & \textbf{1.1806}  \\
                        &                       & RMSE $\downarrow$ & 4.9852 & 2.2084 & 2.8524 & 1.6632 & 1.9723 & 1.6178 & 1.9642 & 2.1820 & 1.6037 & \textbf{1.5806}  \\ 
\hline\hline
\multirow{6}{*}{Task3}  & \multirow{2}{*}{80\%} & MAE $\downarrow$  & 4.6152 & 1.9660 & 2.2356 & 1.6665 & 1.7040 & 1.3959 & 1.5301 & 1.6249 & 1.2703 & \textbf{1.2615}  \\
                        &                       & RMSE $\downarrow$ & 5.3806 & 2.5648 & 2.6043 & 2.0023 & 2.0034 & 1.7816 & 1.7773 & 2.0195 & 1.6804 & \textbf{1.6643}  \\
                        & \multirow{2}{*}{50\%} & MAE $\downarrow$  & 4.5263 & 2.3968 & 2.6720 & 1.5146 & 1.8695 & 1.6962 & 1.6923 & 1.9271 & 1.4362 & \textbf{1.4239}  \\
                        &                       & RMSE $\downarrow$ & 5.3149 & 3.0043 & 3.0200 & 2.2492 & 2.1780 & 2.1265 & 2.0852 & 2.4460 & 1.9404 & \textbf{1.9212}  \\
                        & \multirow{2}{*}{20\%} & MAE $\downarrow$  & 4.6544 & 2.6637 & 2.9198 & 2.2774 & 2.1097 & 1.9358 & 1.9999 & 2.4382 & 1.6272 & \textbf{1.6042}  \\
                        &                       & RMSE $\downarrow$ & 5.4313 & 3.3503 & 3.2551 & 2.6253 & 2.4494 & 2.4132 & 2.3214 & 3.0680 & 2.2106 & \textbf{2.2045}  \\ 
\hline\hline
\multirow{6}{*}{Task4}  & \multirow{2}{*}{80\%} & MAE $\downarrow$  & 4.3608 & 2.3455 & 2.6931 & 2.4435 & 2.2801 & 2.2723 & 1.9876 & 2.2032 & 1.6914 & \textbf{1.6769}  \\
                        &                       & RMSE $\downarrow$ & 5.1976 & 3.1429 & 3.1100 & 3.0077 & 2.7748 & 2.8064 & 2.4001 & 2.7317 & 2.3176 & \textbf{2.3065}  \\
                        & \multirow{2}{*}{50\%} & MAE $\downarrow$  & 4.2722 & 2.5414 & 2.9099 & 2.6365 & 2.5499 & 2.4033 & 2.0094 & 2.8873 & 1.9729 & \textbf{1.9095}  \\
                        &                       & RMSE $\downarrow$ & 5.0742 & 3.3250 & 3.4155 & 3.0877 & 3.0822 & 2.9383 & 2.8739 & 3.4064 & 2.5599 & \textbf{2.5310}  \\
                        & \multirow{2}{*}{20\%} & MAE $\downarrow$  & 4.2999 & 3.0902 & 3.2640 & 3.4904 & 3.3885 & 3.2540 & 2.6021 & 2.9873 & 2.8191 & \textbf{2.4879}  \\
                        &                       & RMSE $\downarrow$ & 5.1121 & 3.8110 & 3.8148 & 3.7434 & 3.6213 & 3.4689 & 3.3458 & 3.2064 & 3.2046 & \textbf{3.0080}  \\ 
\hline\hline
\multirow{6}{*}{Task5}  & \multirow{2}{*}{80\%} & MAE $\downarrow$  & 4.3225 & 2.5785 & 2.1394 & 2.3045 & 2.1690 & 2.2496 & 2.3459 & 2.8020 & 1.9116 & \textbf{1.8794}  \\
                        &                       & RMSE $\downarrow$ & 5.1000 & 3.3743 & 2.7358 & 2.8589 & 2.5646 & 2.5602 & 2.8882 & 3.3021 & 2.5670 & \textbf{2.5496}  \\
                        & \multirow{2}{*}{50\%} & MAE $\downarrow$  & 4.3590 & 2.7830 & 4.0070 & 2.6926 & 2.7990 & 3.0579 & 2.5764 & 3.0005 & 2.1192 & \textbf{2.0534}  \\
                        &                       & RMSE $\downarrow$ & 5.1262 & 3.5396 & 5.1296 & 3.1398 & 3.1982 & 2.9516 & 2.8743 & 3.3544 & 2.7271 & \textbf{2.6920}  \\
                        & \multirow{2}{*}{20\%} & MAE $\downarrow$  & 4.3877 & 3.2192 & 4.2272 & 3.5168 & 3.6585 & 3.3226 & 3.0329 & 3.4071 & 2.8990 & \textbf{2.5692}  \\
                        &                       & RMSE $\downarrow$ & 5.1655 & 3.9194 & 5.3115 & 3.7645 & 3.8855 & 3.4409 & 3.4444 & 3.6752 & 3.3057 & \textbf{3.1255}  \\ 
\hline\hline
\multirow{6}{*}{Task6}  & \multirow{2}{*}{80\%} & MAE $\downarrow$  & 3.7500 & 1.2582 & 2.6992 & 1.1622 & 0.9767 & 0.8511 & 0.9847 & 1.5454 & 0.7953 & \textbf{0.7157}  \\
                        &                       & RMSE $\downarrow$ & 4.5154 & 1.5972 & 3.0487 & 1.4008 & 1.1999 & 1.0859 & 1.2109 & 1.8295 & 1.0114 & \textbf{0.9111}  \\
                        & \multirow{2}{*}{50\%} & MAE $\downarrow$  & 3.7043 & 1.4424 & 2.7474 & 1.5522 & 1.3471 & 0.9391 & 1.1132 & 1.6469 & 0.9143 & \textbf{0.9004}  \\
                        &                       & RMSE $\downarrow$ & 4.4764 & 1.7965 & 3.1459 & 1.8100 & 1.2725 & 1.2059 & 1.4963 & 1.9329 & 1.1721 & \textbf{1.0764}  \\
                        & \multirow{2}{*}{20\%} & MAE $\downarrow$  & 3.7490 & 1.7628 & 3.1040 & 1.8315 & 1.5033 & 1.0688 & 1.1984 & 1.7072 & 0.9373 & \textbf{0.8165}  \\
                        &                       & RMSE $\downarrow$ & 4.5649 & 2.1869 & 3.5774 & 2.0999 & 1.7550 & 1.3725 & 1.5302 & 2.0039 & 1.2025 & \textbf{1.0379}  \\
\bottomrule
\end{tabular}
\end{table*}

\begin{table*}
\centering
\caption{Performance comparison: top-k ranking}
\label{comp:top-k}
\begin{tabular}{ccccccccccccc} 
\toprule
Task                    & Train-ratio           & Metric                    & TGT    & CMF    & CST    & EMCDR  & DCDCSR & SSCDR  & TMCDR  & LACDR  & PTUCDR & CVPM             \\ 
\hline
\multirow{6}{*}{Task-1} & \multirow{2}{*}{80\%} & Hit $\uparrow$            & 0.2081 & 0.4272 & 0.4040 & 0.4226 & 0.4285 & 0.4429 & 0.4317 & 0.4210 & 0.4421 & \textbf{0.4612}  \\
                        &                       & NDCG $\uparrow$           & 0.1112 & 0.2326 & 0.2309 & 0.2215 & 0.2379 & 0.2439 & 0.2334 & 0.2496 & 0.2489 & \textbf{0.2685}  \\
                        & \multirow{2}{*}{50\%} & Hit $\uparrow$            & 0.1805 & 0.3816 & 0.3838 & 0.3513 & 0.4017 & 0.4026 & 0.4088 & 0.4167 & 0.4131 & \textbf{0.4285}  \\
                        &                       & NDCG $\uparrow$           & 0.0938 & 0.2190 & 0.2149 & 0.1869 & 0.2269 & 0.2249 & 0.2154 & 0.2262 & 0.2362 & \textbf{0.2511}  \\
                        & \multirow{2}{*}{20\%} & Hit $\uparrow$            & 0.1808 & 0.3799 & 0.3824 & 0.3416 & 0.3803 & 0.3958 & 0.3747 & 0.3906 & 0.3979 & \textbf{0.4108}  \\
                        &                       & NDCG $\uparrow$           & 0.0963 & 0.2168 & 0.2185 & 0.1632 & 0.2124 & 0.2197 & 0.2100 & 0.2235 & 0.2231 & \textbf{0.2402}  \\ 
\hline\hline
\multirow{6}{*}{Task-2} & \multirow{2}{*}{80\%} & Hit $\uparrow$            & 0.2838 & 0.5520 & 0.5408 & 0.5730 & 0.5665 & 0.5696 & 0.5688 & 0.5523 & 0.5782 & \textbf{0.5810}  \\
                        &                       & NDCG $\uparrow$           & 0.1768 & 0.3608 & 0.3289 & 0.3870 & 0.3758 & 0.3761 & 0.3711 & 0.3494 & 0.3922 & \textbf{0.3930}  \\
                        & \multirow{2}{*}{50\%} & Hit $\uparrow$            & 0.2576 & 0.5350 & 0.5373 & 0.5469 & 0.5565 & 0.5446 & 0.5336 & 0.5433 & 0.5587 & \textbf{0.5652}  \\
                        &                       & NDCG $\uparrow$           & 0.1592 & 0.3243 & 0.3297 & 0.3596 & 0.3698 & 0.3550 & 0.3474 & 0.3433 & 0.3854 & \textbf{0.3868}  \\
                        & \multirow{2}{*}{20\%} & Hit $\uparrow$            & 0.2553 & 0.5062 & 0.5315 & 0.5425 & 0.5500 & 0.5492 & 0.5337 & 0.5314 & 0.5565 & \textbf{0.5591}  \\
                        &                       & NDCG $\uparrow$           & 0.1557 & 0.3058 & 0.3263 & 0.3559 & 0.3660 & 0.3518 & 0.3555 & 0.3375 & 0.3694 & \textbf{0.3702}  \\ 
\hline\hline
\multirow{6}{*}{Task3}  & \multirow{2}{*}{80\%} & Hit $\uparrow$            & 0.1980 & 0.4029 & 0.4128 & 0.4073 & 0.3889 & 0.4001 & 0.3905 & 0.4128 & 0.4128 & \textbf{0.4299}  \\
                        &                       & NDCG $\uparrow$           & 0.1023 & 0.2246 & 0.2417 & 0.2299 & 0.2133 & 0.2157 & 0.2204 & 0.2411 & 0.2324 & \textbf{0.2506}  \\
                        & \multirow{2}{*}{50\%} & Hit $\uparrow$            & 0.2031 & 0.3886 & 0.4058 & 0.3951 & 0.3873 & 0.4198 & 0.4011 & 0.4028 & 0.3957 & \textbf{0.4218}  \\
                        &                       & NDCG $\uparrow$           & 0.1433 & 0.2018 & 0.2370 & 0.1971 & 0.2124 & 0.2303 & 0.2106 & 0.2288 & 0.2204 & \textbf{0.2471}  \\
                        & \multirow{2}{*}{20\%} & Hit $\uparrow$            & 0.1862 & 0.3292 & 0.3597 & 0.3621 & 0.3665 & 0.3782 & 0.3773 & 0.3471 & 0.3686 & \textbf{0.3910}  \\
                        &                       & NDCG $\uparrow$           & 0.0974 & 0.1893 & 0.2022 & 0.2027 & 0.2017 & 0.2127 & 0.2164 & 0.1910 & 0.2024 & \textbf{0.2215}  \\ 
\hline\hline
\multirow{6}{*}{Task4}  & \multirow{2}{*}{80\%} & Hit $\uparrow$            & 0.2447 & 0.5651 & 0.5286 & 0.5261 & 0.5677 & 0.5364 & 0.5338 & 0.5677 & 0.5598 & \textbf{0.5703}  \\
                        &                       & NDCG $\uparrow$           & 0.1583 & 0.3780 & 0.3483 & 0.3458 & 0.3944 & 0.3880 & 0.3877 & 0.3564 & 0.3943 & \textbf{0.4003}  \\
                        & \multirow{2}{*}{50\%} & Hit $\uparrow$            & 0.2031 & 0.4321 & 0.4589 & 0.4130 & 0.4384 & 0.4287 & 0.4163 & 0.4482 & 0.4404 & \textbf{0.4521}  \\
                        &                       & NDCG $\uparrow$           & 0.1497 & 0.3324 & 0.3143 & 0.3136 & 0.3352 & 0.3215 & 0.3201 & 0.3220 & 0.3318 & \textbf{0.3400}  \\
                        & \multirow{2}{*}{20\%} & Hit $\uparrow$            & 0.1923 & 0.3389 & 0.3491 & 0.3462 & 0.3305 & 0.3409 & 0.3421 & 0.3485 & 0.3269 & \textbf{0.3579}  \\
                        &                       & NDCG $\uparrow$           & 0.1320 & 0.2801 & 0.2682 & 0.2741 & 0.2737 & 0.2871 & 0.2781 & 0.2767 & 0.2720 & \textbf{0.2872}  \\ 
\hline\hline
\multirow{6}{*}{Task5}  & \multirow{2}{*}{80\%} & Hit $\uparrow$            & 0.2695 & 0.5112 & 0.4970 & 0.4923 & 0.4960 & 0.4921 & 0.4988 & 0.5117 & 0.5039 & \textbf{0.5195}  \\
                        &                       & NDCG $\uparrow$           & 0.1726 & 0.3340 & 0.3209 & 0.3163 & 0.3547 & 0.3495 & 0.3478 & 0.3365 & 0.3537 & \textbf{0.3619}  \\
                        & \multirow{2}{*}{50\%} & Hit $\uparrow$            & 0.2031 & 0.4319 & 0.4397 & 0.4016 & 0.4140 & 0.4317 & 0.4321 & 0.4375 & 0.4330 & \textbf{0.4441}  \\
                        &                       & NDCG $\uparrow$           & 0.1443 & 0.3050 & 0.2864 & 0.2991 & 0.3057 & 0.2917 & 0.3001 & 0.3041 & 0.3089 & \textbf{0.3130}  \\
                        & \multirow{2}{*}{20\%} & Hit $\uparrow$            & 0.1711 & 0.2670 & 0.2567 & 0.2840 & 0.2862 & 0.2904 & 0.2813 & 0.2911 & 0.2727 & \textbf{0.3034}  \\
                        &                       & NDCG $\uparrow$           & 0.1052 & 0.2110 & 0.2015 & 0.1809 & 0.2142 & 0.2090 & 0.2015 & 0.2139 & 0.2048 & \textbf{0.2157}  \\ 
\hline\hline
\multirow{6}{*}{Task6}  & \multirow{2}{*}{80\%} & Hit $\uparrow$ & 0.1846 & 0.5258 & 0.5315 & 0.5103 & 0.5284 & 0.5330 & 0.5353 & 0.5223 & 0.5417 & \textbf{0.5523}  \\
                        &                       & NDCG $\uparrow$           & 0.0882 & 0.2637 & 0.2404 & 0.2402 & 0.2806 & 0.2758 & 0.3173 & 0.2705 & 0.3214 & \textbf{0.3219}  \\
                        & \multirow{2}{*}{50\%} & Hit $\uparrow$            & 0.1323 & 0.4911 & 0.4970 & 0.5029 & 0.4594 & 0.5247 & 0.5247 & 0.4735 & 0.5370 & \textbf{0.5400}  \\
                        &                       & NDCG $\uparrow$           & 0.0618 & 0.2444 & 0.2533 & 0.2606 & 0.2805 & 0.2593 & 0.2580 & 0.2988 & 0.2896 & \textbf{0.3064}  \\
                        & \multirow{2}{*}{20\%} & Hit $\uparrow$            & 0.0981 & 0.3925 & 0.3001 & 0.4032 & 0.4000 & 0.4862 & 0.4525 & 0.4255 & 0.4922 & \textbf{0.5111}  \\
                        &                       & NDCG $\uparrow$           & 0.0442 & 0.1992 & 0.1517 & 0.2061 & 0.2295 & 0.2792 & 0.2378 & 0.2190 & 0.2672 & \textbf{0.2822}  \\
\bottomrule
\end{tabular}
\end{table*}
\noindent\textbf{Comparison of different overlapping user ratios:}
We also show the experimental results under different data set partitions in Table~\ref{comp:value}-\ref{comp:top-k}. When there are fewer overlapping users, collaborative filtering signals decrease and the training of transfer patterns becomes incomplete, which presents greater challenges.
This can be observed from the performance decline of all algorithms to varying degrees as the proportion decreases.
EMCDR and DCDCSR exhibit significant fluctuations in performance at different ratios, with a noticeable drop at low ratios, even approaching the CMF of mixed training. This indicates their sensitivity to the proportion of overlapping users and their limited applicability in sparse overlapping user scenarios.
Benefiting from meta-learning for generalization to new tasks, even without introducing explicit supervision signals for non-overlapping user knowledge, PTUCDR still achieves good ranking and valuation capabilities in 50\% of scenarios, but this ability dramatically deteriorates at a 20\% overlap rate. SSCDR and \textit{CVPM} exhibit greater robustness, as they both utilize self-supervised learning to extract knowledge from non-overlapping users. Despite minor fluctuations, SSCDR's low performance in each task stems from the common mapping function and the mutual inference of mapping targets.

In general, we employ the meta-network to enhance the generalization of the model to new users and leverage the self-supervised signals of non-overlapping users to get rid of the dependence on overlapping users, thus having a wider application scenario.

\noindent\textbf{Comparison of cross-domain and cross-system tasks:}
Cross-domain and cross-system tasks have partial overlap in terms of user-side and item-side, respectively. For the latter, we align the data based on the movie name in task 6.
As can be seen from Table~\ref{comp:value}-\ref{comp:top-k}, \textit{CVPM} exhibits excellent valuation and ranking capabilities in both scenarios.
Both CMF and CST have consistently performed poorly in cross-domain and cross-system recommendation tasks, as their coarse-grained transfer strategies are insufficient to meet the challenges posed by complex cross-domain scenarios.
EMCDR and DCDCSR exhibit poor performance on cross-system recommendation, in contrast to their performance on cross-domain tasks. Our argument is that it is challenging for a common mapping function to learn correlations between items, given the significant variability of item features and limited overlap.
TMCDR and PTUCDR show no marked change in performance, gaining from the good generalization of the meta-learning pipeline on new tasks, whether user or item.

Overall, \textit{CVPM} can achieve superior performance in any scenario, and is particularly well-suited for challenging tasks that involve large domain differences, such as cross-system recommendation.

\section{Model Analysis and Discussion}\label{section6}
In this section, to further demonstrate the superiority of our proposed framework, we conduct extensive experimental validation from \textit{feature visualization, underlying model replacement}, and \textit{ablation experiments}, etc.

\begin{figure}[!ht] \centering   
\subfigure[Task 1-MAE] { 
\label{warm:1mae}     
\includegraphics[width=0.45\linewidth,height=0.4\linewidth]{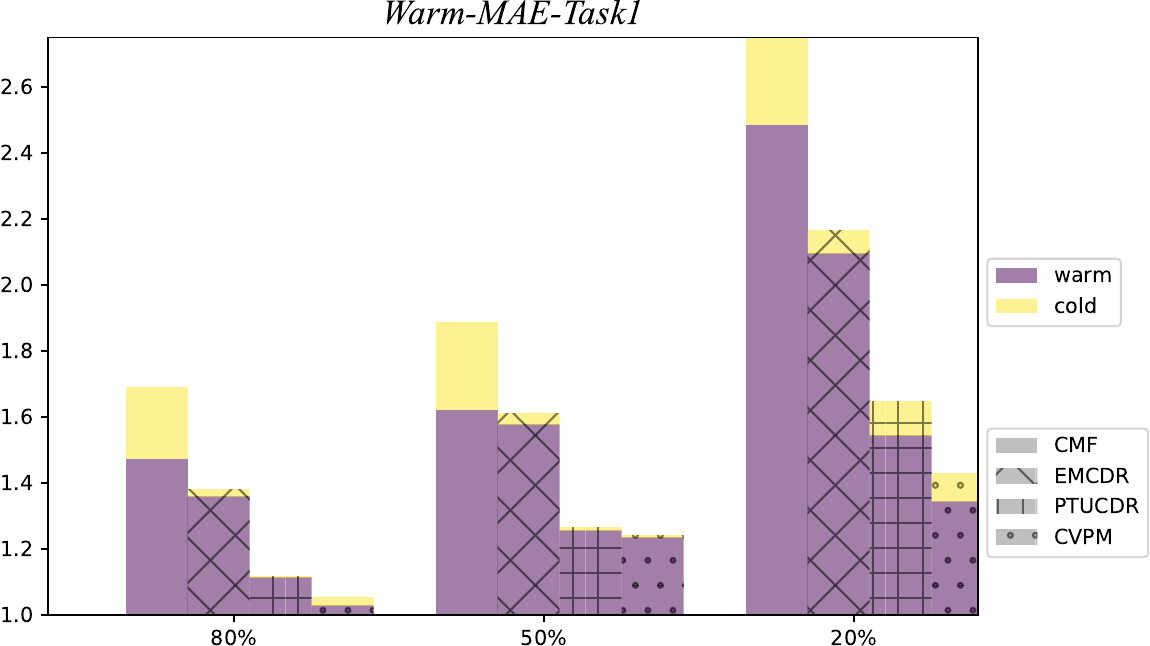}
}    
\subfigure[Task 1-NDCG] { 
\label{warm:1ndcg}     
\includegraphics[width=0.45\linewidth,height=0.4\linewidth]{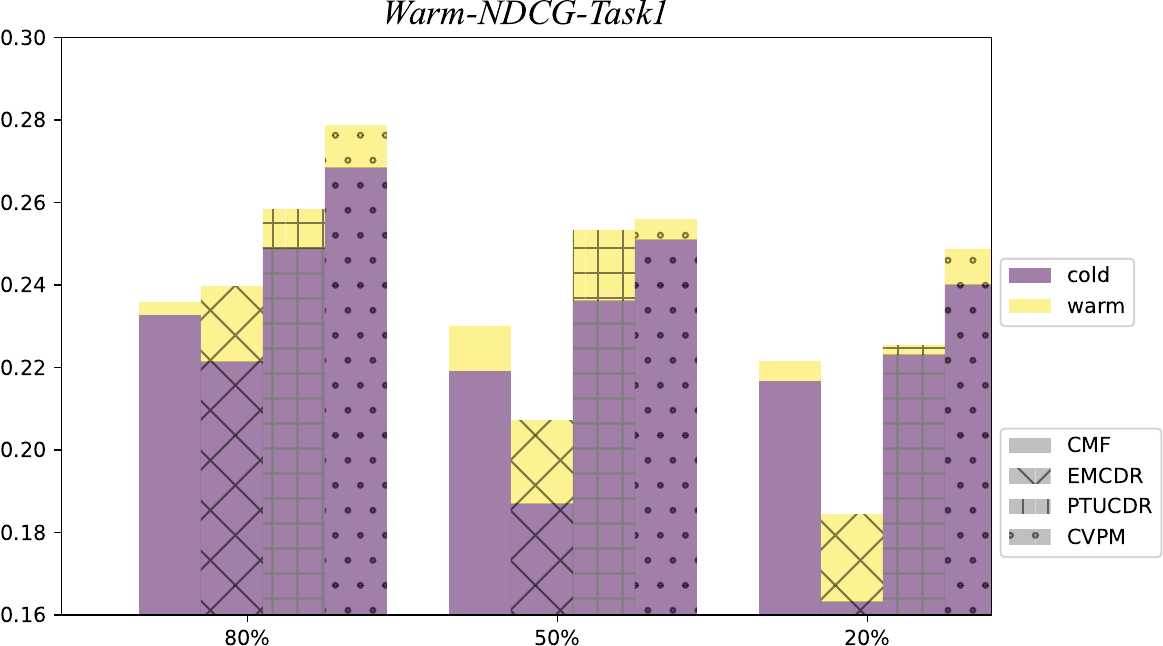}
} 

\subfigure[Task 6-MAE] { 
\label{warm:6ndcg}     
\includegraphics[width=0.45\linewidth,height=0.4\linewidth]{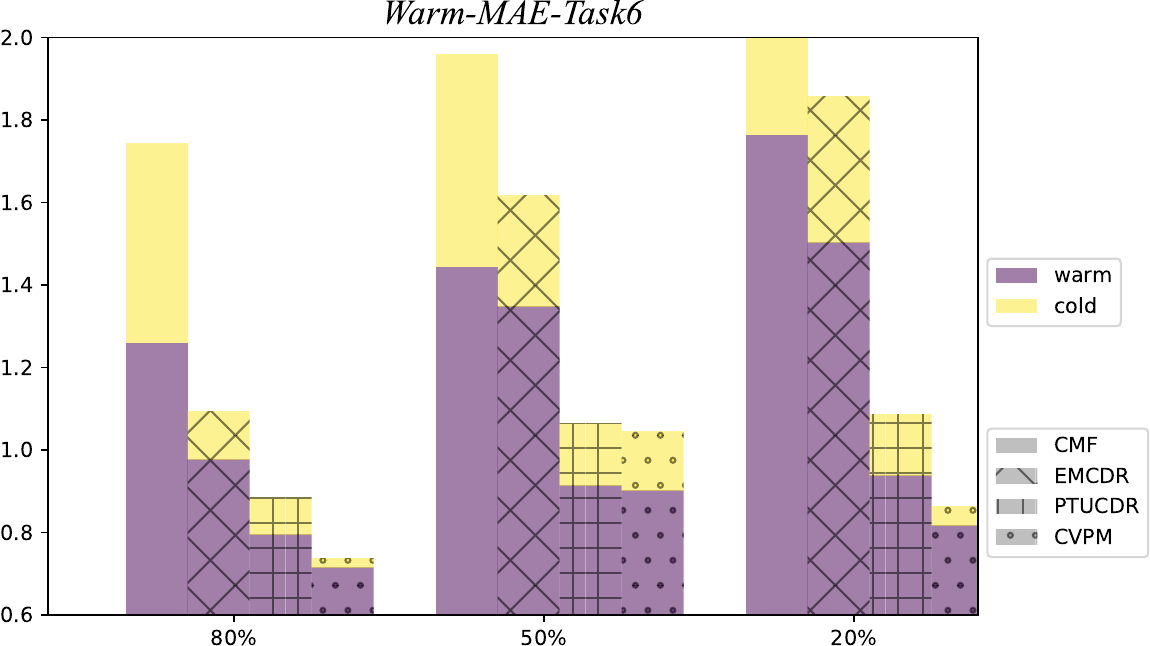}
}    
\subfigure[Task 6-NDCG] { 
\label{warm:6ndcg}     
\includegraphics[width=0.45\linewidth,height=0.4\linewidth]{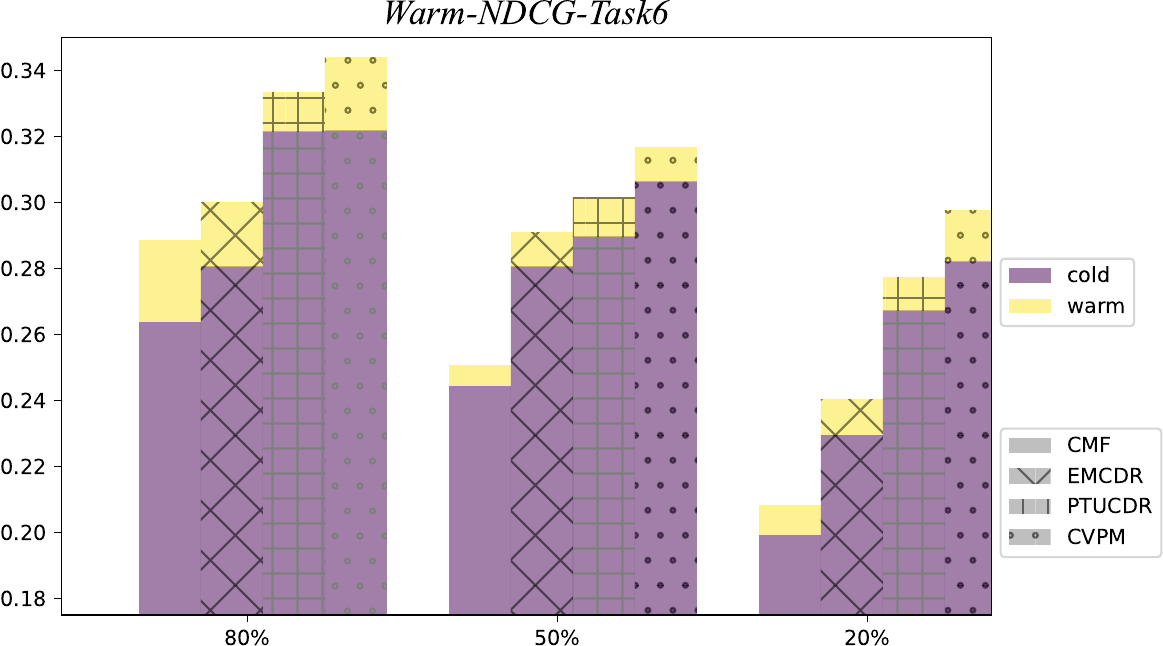}
} 

\centering
\caption{Warm-start experiment.}     
\label{warm}     
\end{figure}
\subsection{Warm-start Recommendation Scenario}
Unlike previous work that only examines model performance on cold-start tasks, we consider another real-world application scenario where user interactions are small but insufficient to capture their preferences, also called \textit{warm-start}. 
In other words, in this scenario, instead of blocking all interactions of the test user in the target domain, we keep the first 50\% of interactions for model fine-tuning and the last 50\% for testing. We choose MAE and NDCG to measure the value estimation and ranking ability of the algorithm respectively. Note that we sort the interaction records chronologically here to avoid label leakage.
As shown in Fig.~\ref{warm}, in the warm-start scenario, all models perform better, even more than 50\% improvement, as the real supervision signal can force the model to be closer to the real distribution during the fine-tuning stage. PTUCDR and CDVM exhibit marginal improvements relative to other algorithms, primarily due to their close proximity to the performance upper bound. Furthermore, their personalized transfer strategy effectively mitigates the reliance on warm-start items in the target domain. It is worth noting that \textit{CVPM} significantly outperforms the contrasting EMCDR and PUTCDR on MAE, demonstrating its strong value estimation capability. In terms of NDCG, \textit{CVPM} also outperforms the strongest baseline by 8\% in cross-domain recommendation (task 1) and 3\% in cross-system recommendation (task 6), which also proves that it is able to identify users' true preferences.
In general, CVPM has stronger adaptability to cross-domain knowledge transfer in different scenarios.

\subsection{Top-K Performance}
\begin{figure}[!ht] \centering   
\subfigure[Task 1-Hit] { 
\label{topk:1h}     
\includegraphics[width=0.45\linewidth,height=0.40\linewidth]{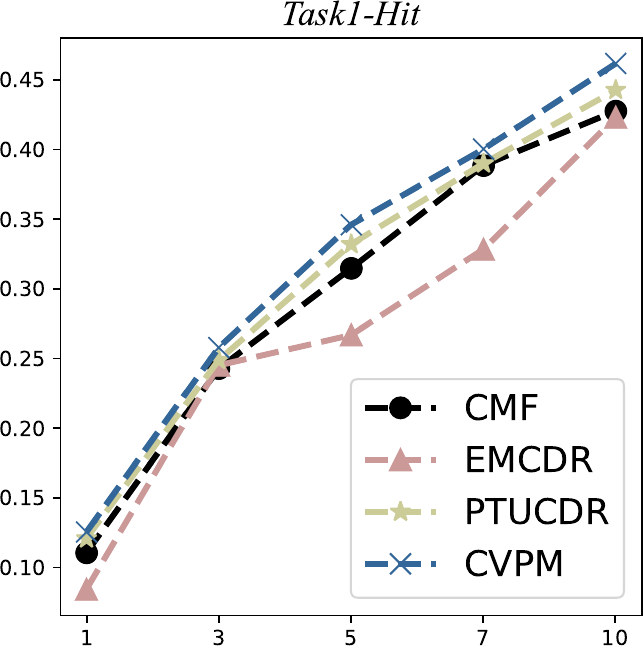}
}    
\subfigure[Task 1-NDCG] { 
\label{topk:1n}     
\includegraphics[width=0.45\linewidth,height=0.40\linewidth]{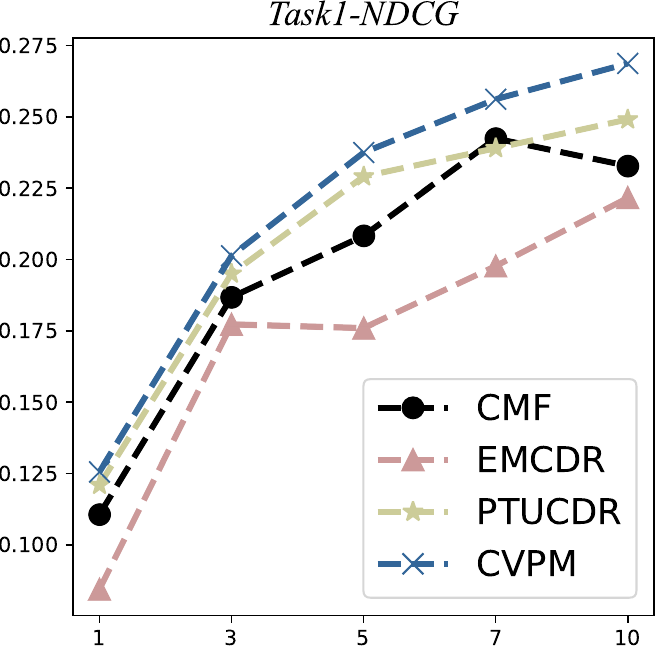}
} 

\subfigure[Task 6-Hit] { 
\label{topk:6h}     
\includegraphics[width=0.45\linewidth,height=0.40\linewidth]{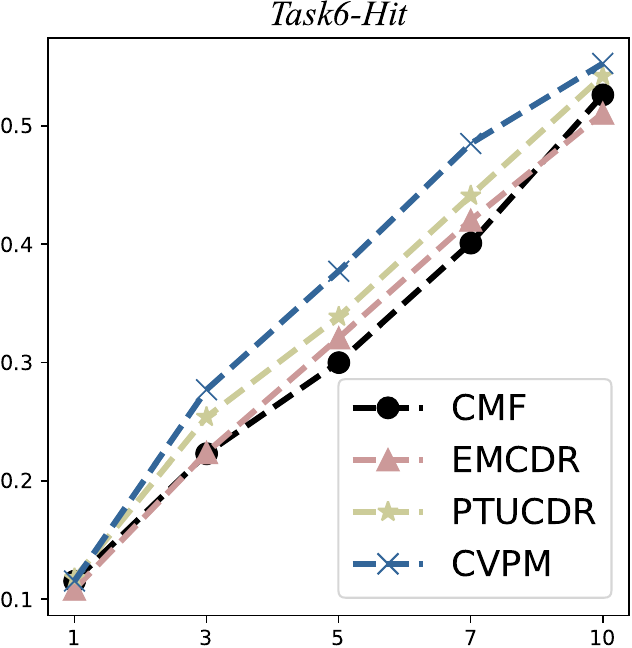}
}    
\subfigure[Task 6-NDCG] { 
\label{topk:6n}     
\includegraphics[width=0.45\linewidth,height=0.40\linewidth]{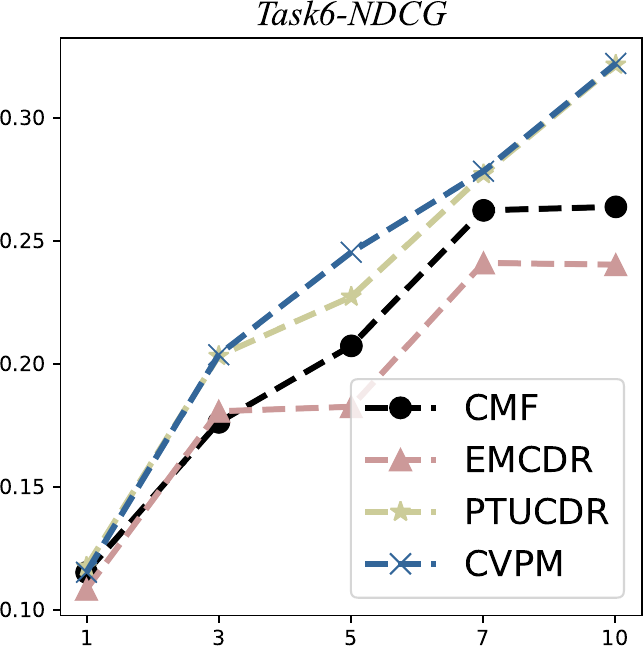}
} 

\centering
\caption{Top-K performance.}     
\label{top-k}     
\end{figure}

The difficulty of recommendation and the length of the recommendation list exhibits an inverse trend, highlighting the importance of exploring the model's performance across different Top-K values to assess its robustness.
Therefore, we assess the ranking performance of several competitive algorithms, including CMT, EMCDR, PTUCDR, and \textit{CVPM}, across recommendation lists of varying lengths. 
As depicted in Figure ~\ref{top-k}, our model consistently outperforms competing baselines across various list lengths, showcasing its exceptional performance and stability.
Specifically, as the value of K increases, the Hit and NDCG metrics of each model in both tasks demonstrate an upward trend, aligning with our expectations. Notably, EMCDR falls significantly behind CMF in the Top-3 and Top-5 scenarios, underscoring the insufficiency of a common mapping function in transferring personalized interests. Even when facing the challenging top-5 difficult ranking scenarios, our model showcases superior ranking capability, surpassing the best baselines by 4\% in both evaluation metrics.

\subsection{T-sne of Entity Represenations}
\begin{figure*}[!ht] \centering   
\subfigure[CMF] { 
\label{tsne:d}     
\includegraphics[width=0.20\linewidth,height=0.20\linewidth]{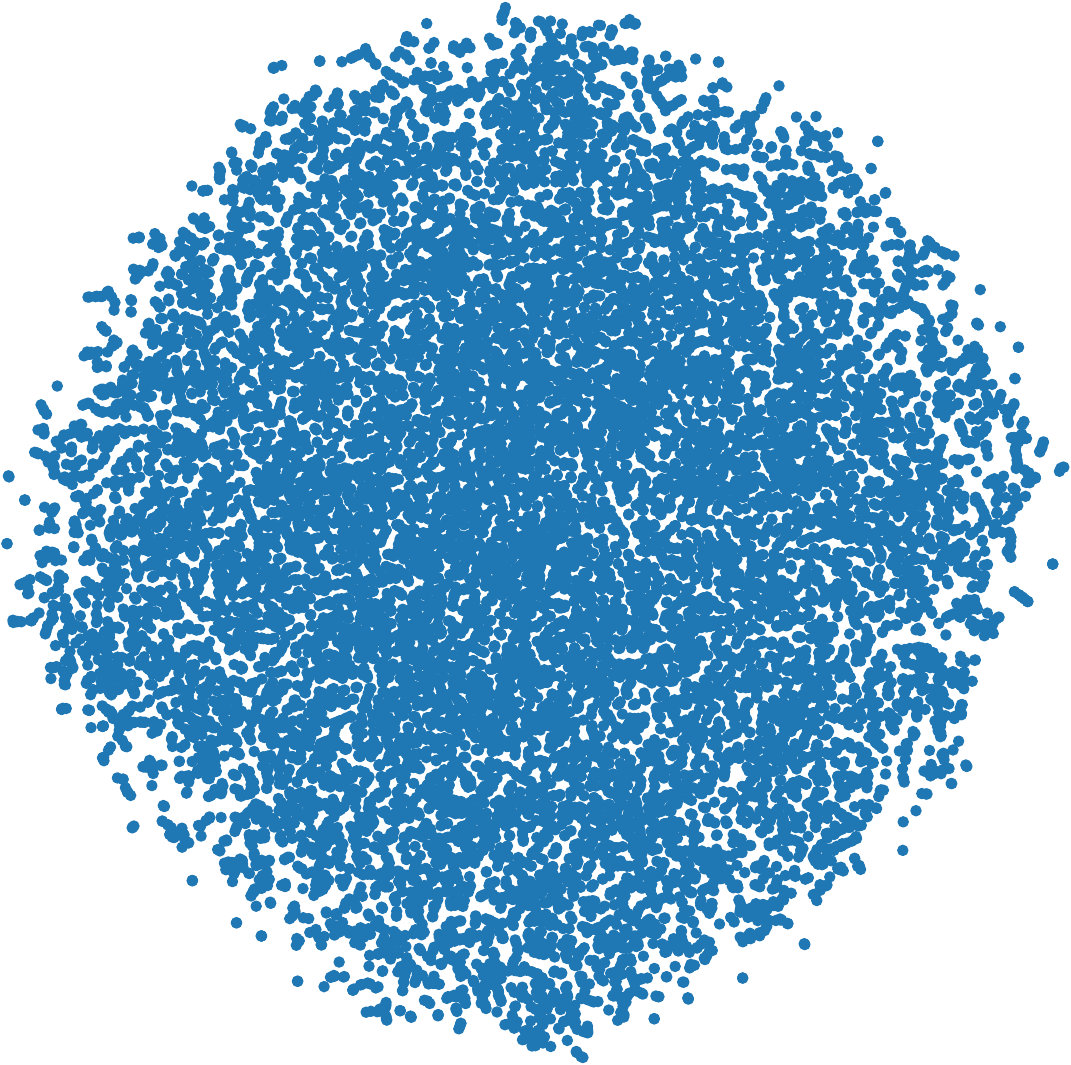}
}    
\subfigure[EMCDR] { 
\label{tsne:e}     
\includegraphics[width=0.20\linewidth,height=0.20\linewidth]{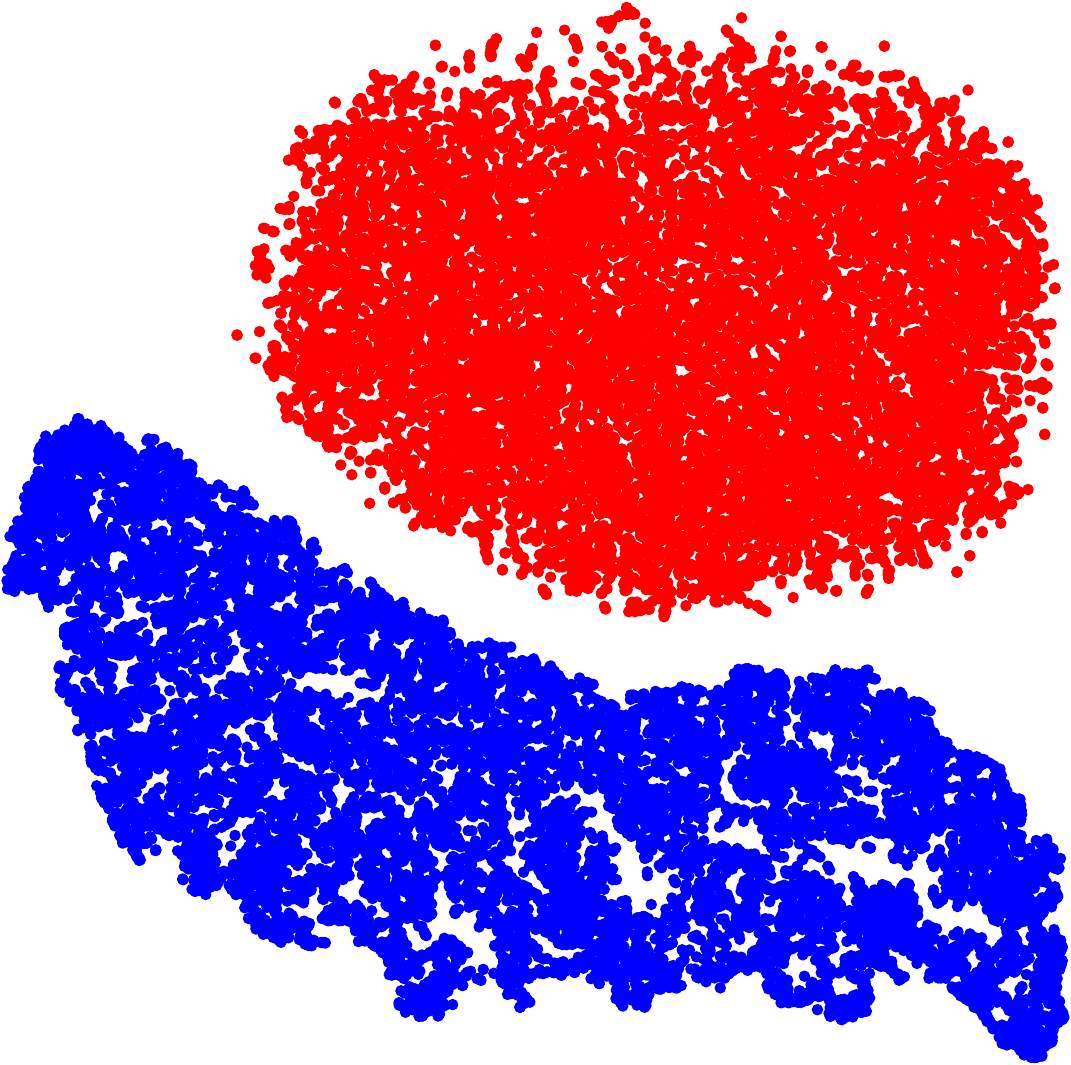}
}
\subfigure[PTUDCDR] { 
\label{tsne:p}     
\includegraphics[width=0.20\linewidth,height=0.20\linewidth]{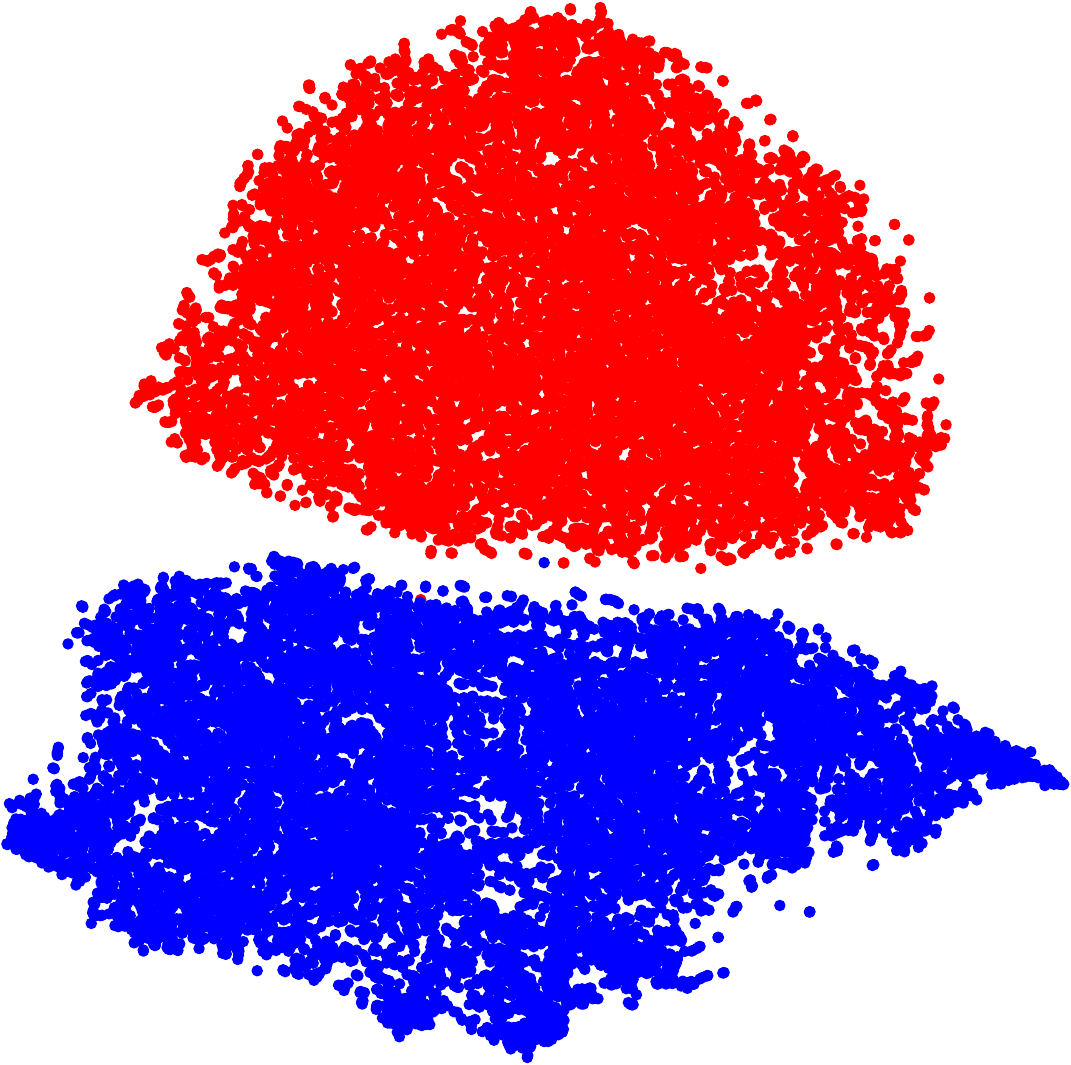}
}    
\subfigure[Ours] { 
\label{tsne:o}     
\includegraphics[width=0.20\linewidth,height=0.20\linewidth]{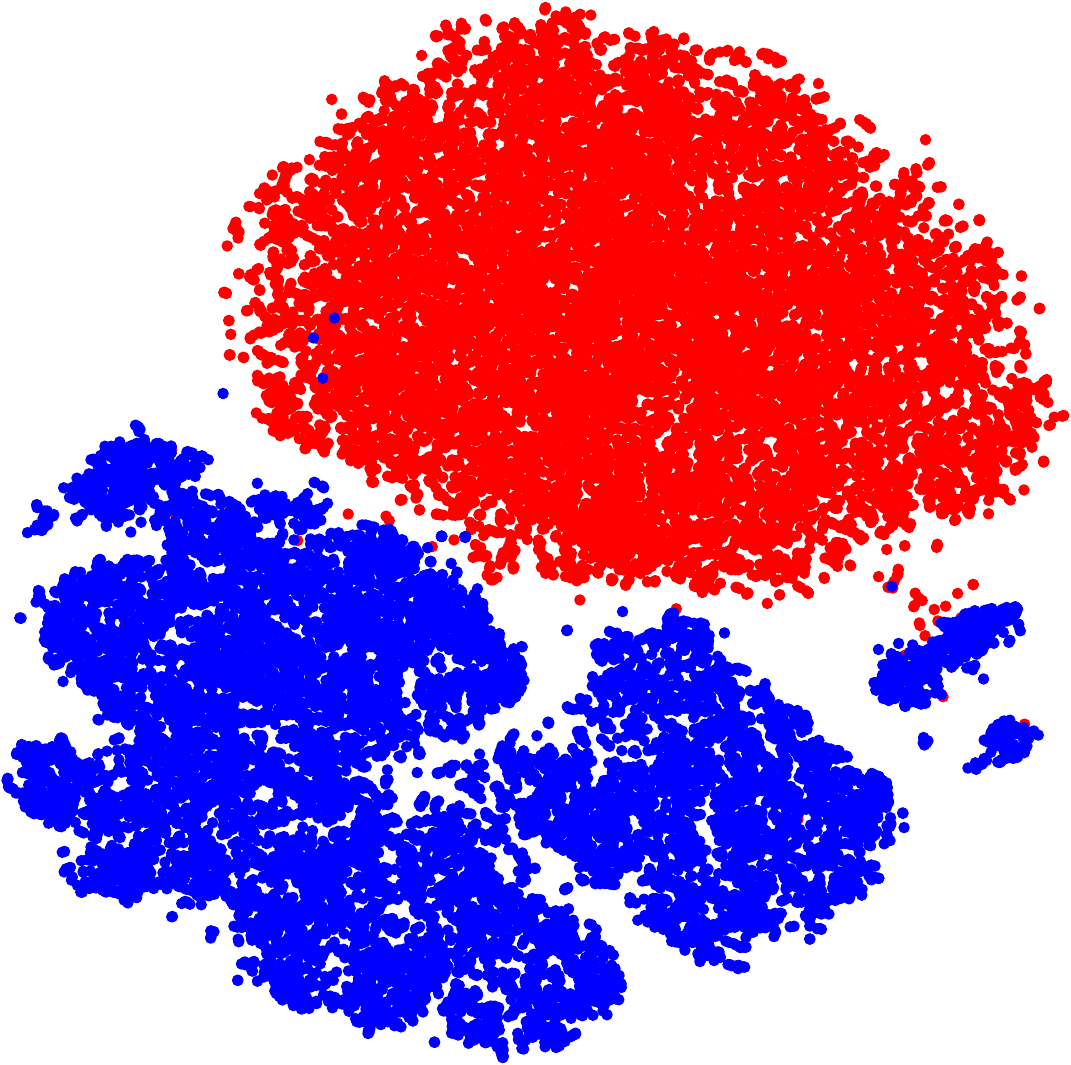}
}
\centering
\caption{T-sne of entity representations. Red represents the source domain representation, and blue represents the mapped target domain representation, where CMFs share the same embedding in both domains.}     
\label{tsne}
\end{figure*}

We visualize sample embeddings before and after transfer using the t-sne technique in task 1 to gain insight into why our model outperforms the baselines. 
As shown in Fig.~\ref{tsne}, representations tend to become linearly similar when mapped by EMCDR, as the common mapping equation disregards the nuances in user interests. 
The representation formed by PTUCDR is more decentralized, benefiting from the introduction of a personalized bridge. However, due to the sparsity of overlapping users and the suboptimal representation of user interests, PTUCDR fails to preserve the semantic structure from the source domain.
In contrast, the CVPM-mapped representations present multiple clusters and are more dispersed.
This not only demonstrates that CVPM is capable of generating more semantic representations, but also preserves the group relations among users.
Overall, EMCDR and PTUCDR pursue common and personalized mapping functions respectively, while \textit{CVPM} considers both.

\subsection{Robustness to the Underlying Model}
\begin{figure*}[!ht] \centering   
\subfigure[GMF-MAE] { 
\label{base:1mae}     
\includegraphics[width=0.20\linewidth,height=0.20\linewidth]{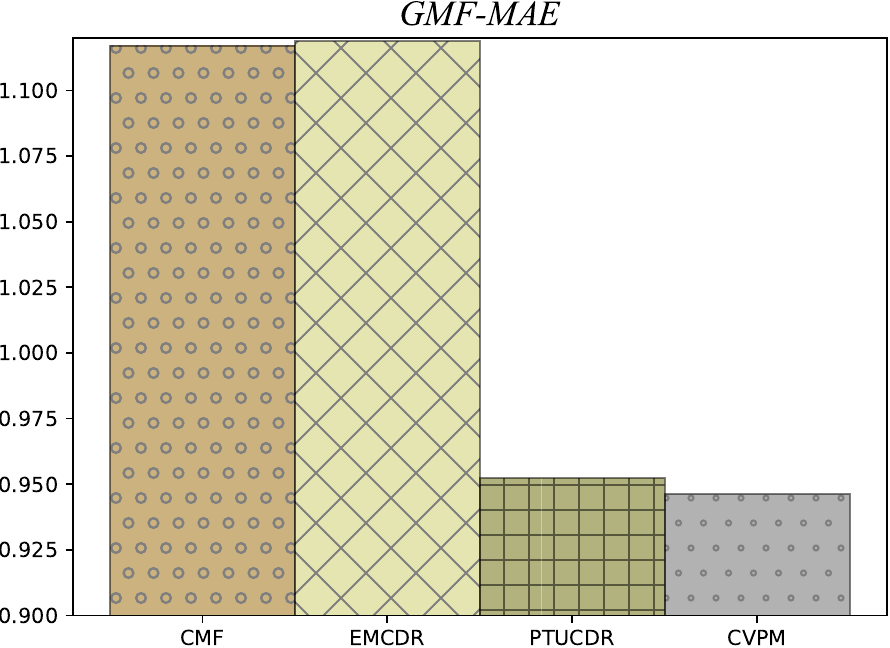}
}    
\subfigure[GMF-NDCG] { 
\label{base:1ndcg}     
\includegraphics[width=0.20\linewidth,height=0.20\linewidth]{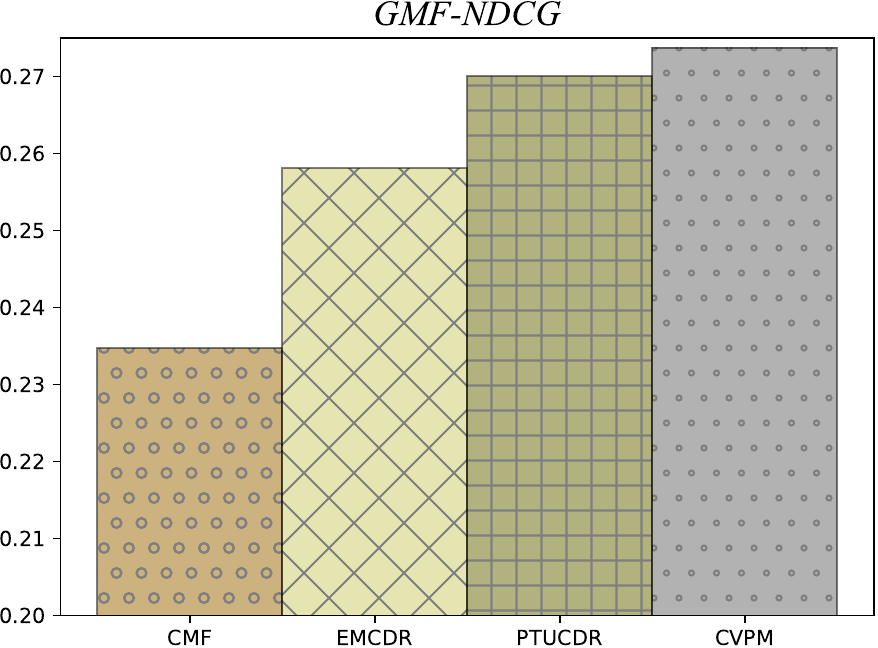}
}
\subfigure[DNN-MAE] { 
\label{base:6mae}     
\includegraphics[width=0.20\linewidth,height=0.20\linewidth]{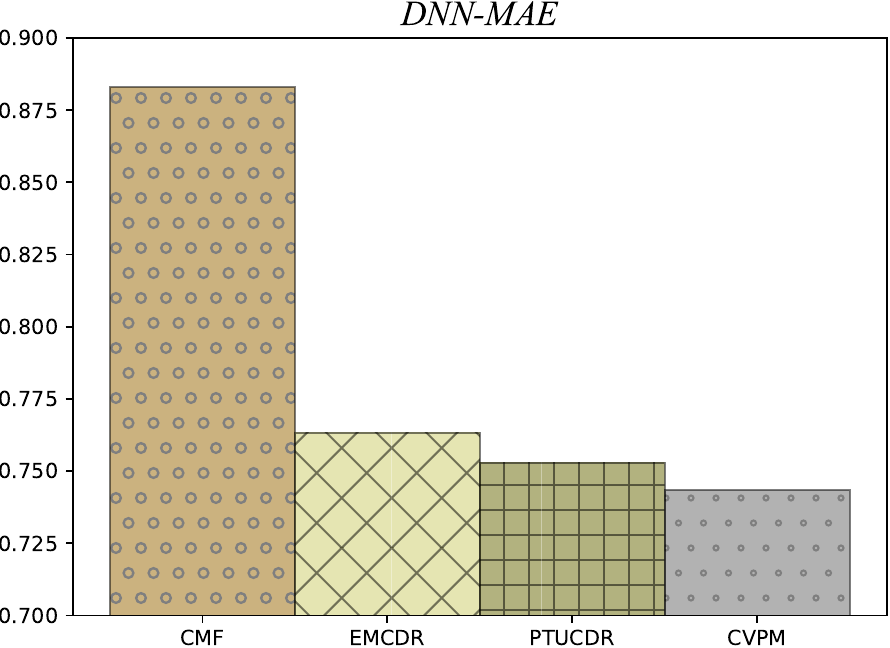}
}    
\subfigure[DNN-NDCG] { 
\label{base:6ndcg}     
\includegraphics[width=0.20\linewidth,height=0.20\linewidth]{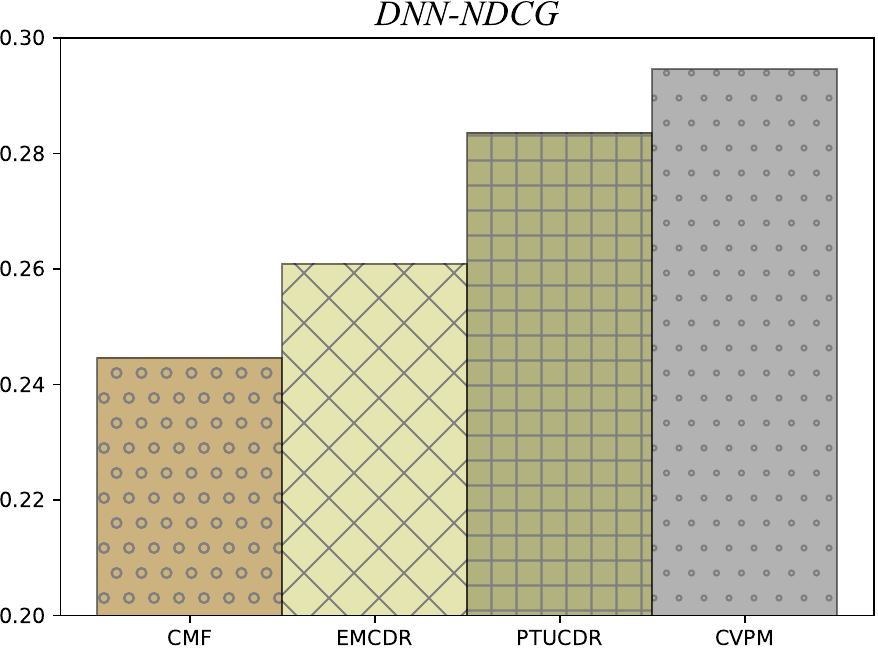}
}
\centering
\caption{Performance under different underlying models.}     
\label{base}
\end{figure*}

One advantage of \textit{CVPM} is that it is model-agnostic, allowing it to be easily replaced by more advanced models. We conduct a series of tests by substituting various underlying models in task 1, including GMF and Youtube DNN~\cite{covington2016deep}, and compare the performance of \textit{CVPM} with that of CMF, EMCDR, and PTUCDR, where we also replace their underlying models. As shown in Fig.~\ref{base}, \textit{CVPM} consistently outperform other models, regardless of the underlying model utilized. Moreover, as the complexity of the underlying model increased, the performance gap between \textit{CVPM} and other models widened. Therefore, \textit{CVPM} is robust to the underlying model and has the potential to improve its performance by incorporating more advanced algorithms. For complex scenes, we recommend using a sophisticated underlying model to further enhance \textit{CVPM}'s performance.

\begin{figure}[!h] \centering   
\subfigure[Ablation-MAE] { 
\label{aba:mae}     
\includegraphics[width=0.45\linewidth,height=0.40\linewidth]{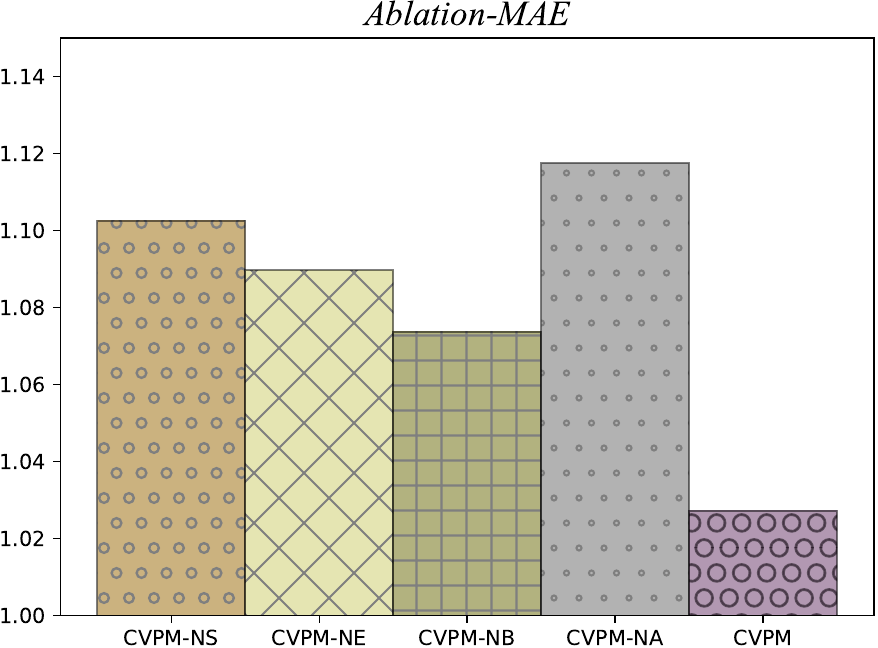}
}    
\subfigure[Ablation-NDCG] { 
\label{aba:ndcg}     
\includegraphics[width=0.45\linewidth,height=0.40\linewidth]{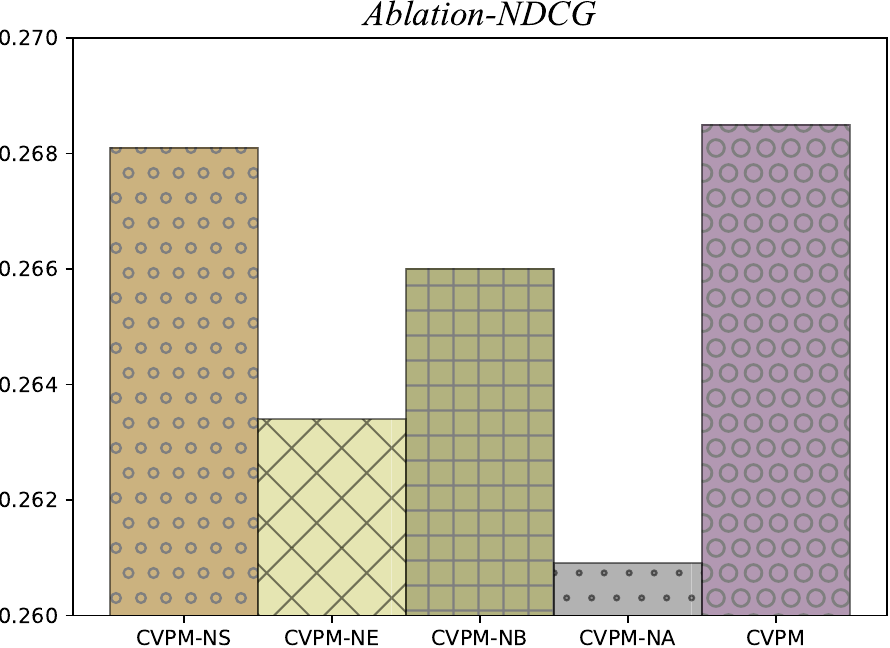}
}    
\centering
\caption{Ablation experiments.}
\label{aba}
\end{figure}

\subsection{Ablation Experiment}
To test the effectiveness of our proposed innovative design, we conduct ablation experiments in task 1 in this subsection.
\begin{itemize}[leftmargin=12pt]
	\item \textbf{\textit{CVPM}-NS:} This variant only utilizes the real interaction data of users when calculating the positive and negative distributions, without sampling process.
	\item \textbf{\textit{CVPM}-NE:} It only employs an encoder to learn user preferences, without fine-grained probabilistic representations to distinguish positive and negative preferences.
	\item \textbf{\textit{CVPM}-NB:} Source-target domain representation transfer using only common mapping, without personalized bias.
	\item \textbf{\textit{CVPM}-NA:} This variant only receives supervision signals from overlapping users, without the self-supervision loss designed for non-overlapping users.
\end{itemize}
As shown in Fig.~\ref{aba}, \textit{CVPM} consistently outperforms any variant on various metrics, proving that any submodule contributes to \textit{CVPM} high performance. The weak performance of CVPM-NB is attributed to its degradation to learning a common mapping function like EMCDR, which conflicts with the personalization purpose of recommender systems. CVPM-NE does not distinguish the heterogeneity of positive and negative preferences at a fine-grained level, resulting in semantically weak and noisy user representations and significant performance drop of 6\%. Compared with \textit{CVPM}, the value estimation and ranking ability of \textit{CVPM}-NA has remarkably declined, mainly because of ignoring non-overlapping user knowledge. CVPM-NS is the variant with the closest performance to \textit{CVPM} on cross-domain tasks, benefiting from it retaining most of \textit{CVPM}'s modular structure. Nonetheless, suffering from limited user interactions, it is incapable of learning comprehensive user preference distribution, leading to suboptimal result on cross-system tasks.
To sum up, any module of \textit{CVPM} plays an indispensable role, demonstrating the advanced nature of our design.

\begin{figure} \centering   
\subfigure[Hyper $d$] { 
\label{hyper:d}     
\includegraphics[width=0.40\linewidth,height=0.35\linewidth]{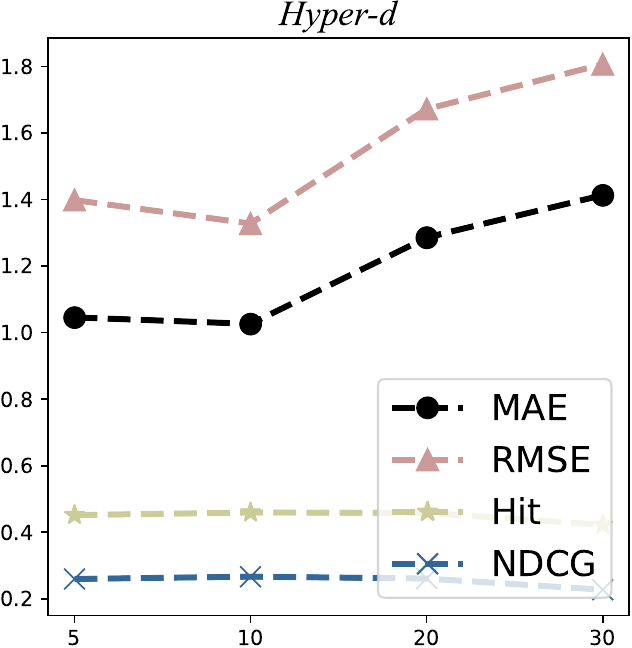}
} 
\subfigure[Hyper $|\mathcal{\ddot{N}}|$] { 
\label{hyper:n}     
\includegraphics[width=0.40\linewidth,height=0.35\linewidth]{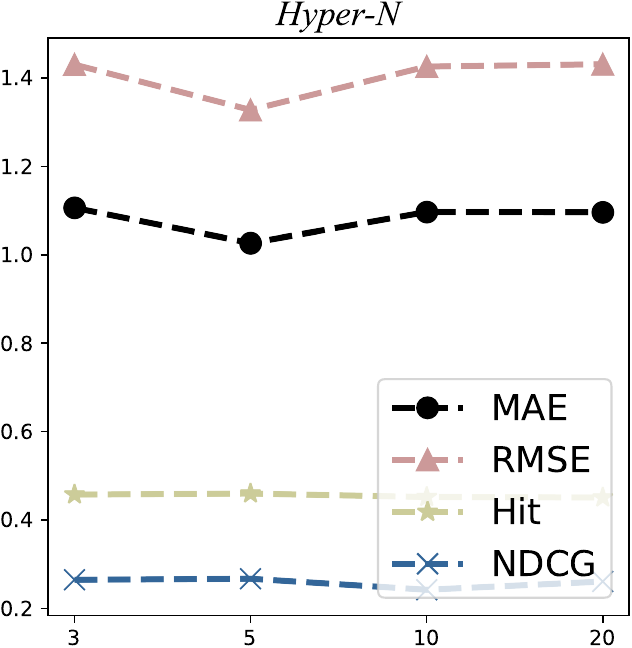}
}
\subfigure[Hyper $|\mathcal{C}|$] { 
\label{hyper:c}     
\includegraphics[width=0.40\linewidth,height=0.35\linewidth]{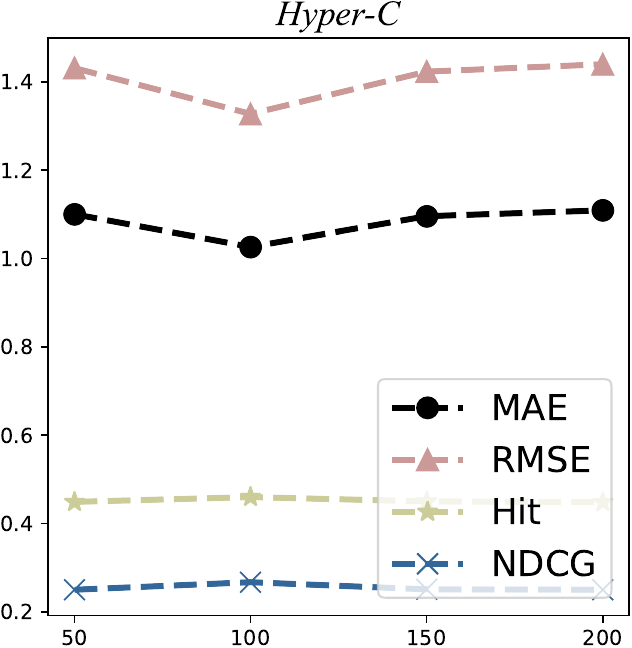}
}    
\subfigure[Hyper $\gamma$] { 
\label{hyper:g}     
\includegraphics[width=0.40\linewidth,height=0.35\linewidth]{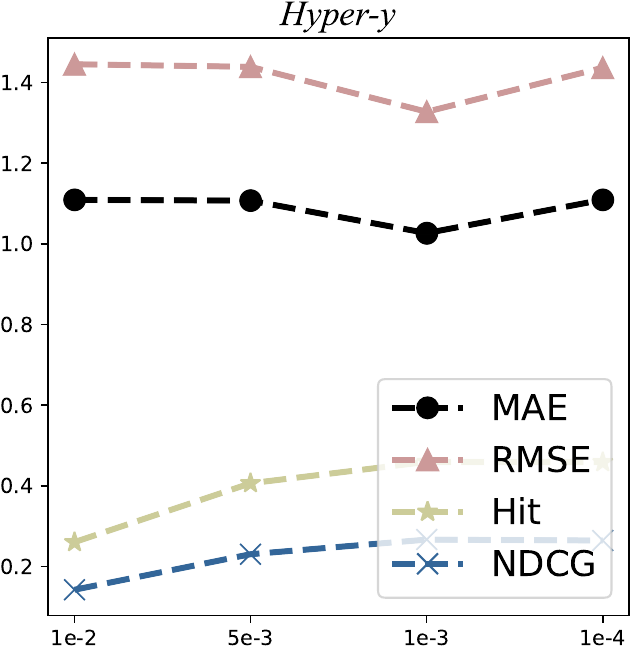}
}
\centering
\caption{Hyper-parameter testing.}     
\label{hyper}

\end{figure}

\subsection{Hyper-parameter Testing}\label{tune}
Next, we conduct several hyperparameter experiments to examine the impact of several key parameters on the \textit{CVPM}.

\noindent\textbf{Embedding Size-$d$}
Embedding size is closely related to the capacity of the model and its ability to extract complex relationships between variables. As shown in Fig.~\ref{hyper:d}, as the embedding size increases, the model improves both in value estimation and top-k ranking ability. \textit{CVPM} achieves the best performance at $d=10$, but starts to decline after that, which suffers from model over-fitting. We recommend using larger embedding size on larger data sets, otherwise set them smaller.

\noindent\textbf{Sampling Size-$|\mathcal{\ddot{N}}|$}
$|\mathcal{\ddot{N}}|$ refers to the number of pseudo positive and negative preference samples used to supplement real interaction data. Larger $|\mathcal{\ddot{N}}|$ will lead to a more complete and semantic preference probability representation, but may introduce some erroneous and non-preference samples. As shown in Fig.~\ref{hyper:n}, our model peaks at $|\mathcal{\ddot{N}}|=5$, proving the need for a trade-off between model semantic completeness and accuracy.

\noindent\textbf{Centroids-$|\mathcal{C}|$}
The number of centroids is a hyperparameter when performing contrastive learning at the group level, representing the granularity of the distribution of interest in the source domain. As the number of centroids increases, the transferred representations can align users' interests from a finer granularity, and vice versa, the coarser granularity. Therefore, an appropriate granularity is very important to preserve user semantics and instance discriminativeness. As shown in Fig.~\ref{hyper:c}, our model achieves the best performance when $|\mathcal{C}|=100$. 

\noindent\textbf{Auxiliary task weight-$\gamma$}
The weights of the self-supervised loss affect the model's ability to distinguish instance preferences, and represent a trade-off between overlapping and non-overlapping user utilization.The evidence in Fig.~\ref{hyper:g} shows that when $\gamma=1e-3$, all metrics achieve better results. With the increase of $\gamma$, the ability of the transferred target domain features to capture group-level semantics and individual-level semantics increases, while the impact of the main supervision loss gradually decreases. Considering that the interaction of overlapping users is the 
primary supervisory signal for constructing the mapping function, the model performance degrades when $\gamma$ is too large.

\section{Conclusion}\label{section7}
In this paper, we have proposed an innovative cross-domain recommendation framework named \textit{CVPM}, which aims to promote the personalization and completeness of interest transfer and minimizing the reliance on overlapping users.
In particular, \textit{CVPM} leverages the valence preference theory to gain insight into user preferences, discriminatively learning users' positive and negative preference distributions.
Relying on meta-learning, we learn a unique bias network for each user and perform interest transfer together with traditional common mapping functions, thus enhancing personalization.
Additionally, \textit{CVPM} incorporates self-supervised signals at both group and individual levels to enhance the semantics of representations and instance discrimination for non-overlapping users.
We conduct extensive experiments on 6 transfer tasks constructed from 8 data sets, including cold-start and warm-start scenarios, representation analysis and ablation experiments, etc., all of which demonstrate the effectiveness and advancement of the proposed framework.

Even though \textit{CVPM} achieves great target domain recommendation performance with the help of rich knowledge in the source domain, our model still has some limitations, which we leave as future work. First,
it is worth exploring how to extend our design and mitigate the negative transfer phenomenon in multi-domain scenarios. Second, how to find the most suitable source domain is still an open problem.

\ifCLASSOPTIONcompsoc
  \section*{Acknowledgments}
\else
  \section*{Acknowledgment}
\fi

This study was partially funded by the supports of National Natural Science Foundation of China (72101176).
\ifCLASSOPTIONcaptionsoff
  \newpage
\fi



%
\bibliographystyle{IEEEtran}

\bibliography{main}


\vspace{-20pt}

\begin{IEEEbiography}[{\includegraphics[width=1in,height=1.25in,clip,keepaspectratio]{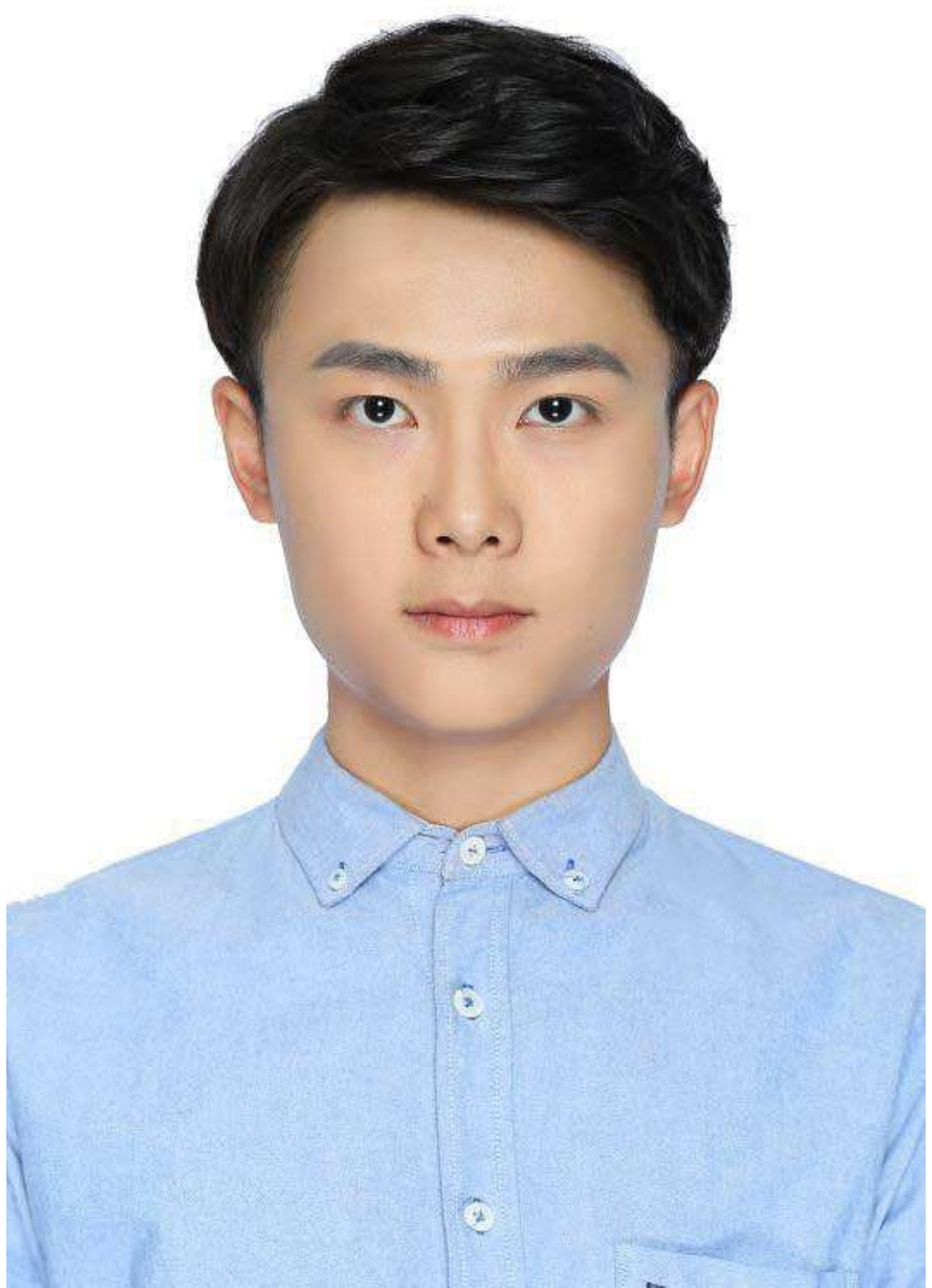}}]{Chuang Zhao} is currently pursuing the Ph.D. degree with The Hong Kong University of Science and Technology, Hong Kong, China. He received his master degree in Tianjin University, Tianjin, China. His research interest includes transfer learning, recommendation system and data mining. He has published some papers in journals and
conference proceedings, including AAAI, WWW, \emph{
INFORMS Journal on Computing} and \emph{IEEE Transactions on Evolutionary Computation}.
\end{IEEEbiography}

\vspace{-20pt}

\begin{IEEEbiography}[{\includegraphics[width=1in,height=1.25in,clip,keepaspectratio]{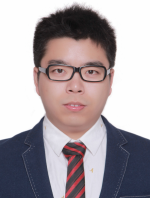}}]
{Hongke Zhao} received the Ph.D. degree from the University of Science and Technology of China (USTC), Hefei, China. He is an associate professor with the College of Management and Economics, Tianjin University. His research interest includes data mining, data-driven management, knowledge and behavior computing.

He has published more than 70 papers in refereed journals and conference proceedings, such as \emph{IEEE Transactions on Knowledge and Data Engineering}, \emph{INFORMS Journal on Computing}, \emph{IEEE Transactions on Evolutionary Computation}, \emph{ACM Transactions on Information Systems}, \emph{ACM Transactions on Intelligent Systems and Technology}, \emph{IEEE Transactions on Systems, Man, and Cybernetics: Systems}, \emph{IEEE Transactions on Big Data}, \emph{Industrial Marketing Management}, \emph{Information Processing \& Management}, \emph{Scientometrics}, ACM SIGKDD, ACM WSDM, ACM SIGIR, IJCAI, AAAI and IEEE ICDM. 
\end{IEEEbiography}

\vspace{-20pt}
\begin{IEEEbiography}[{\includegraphics[width=1in,height=1.25in,clip,keepaspectratio]{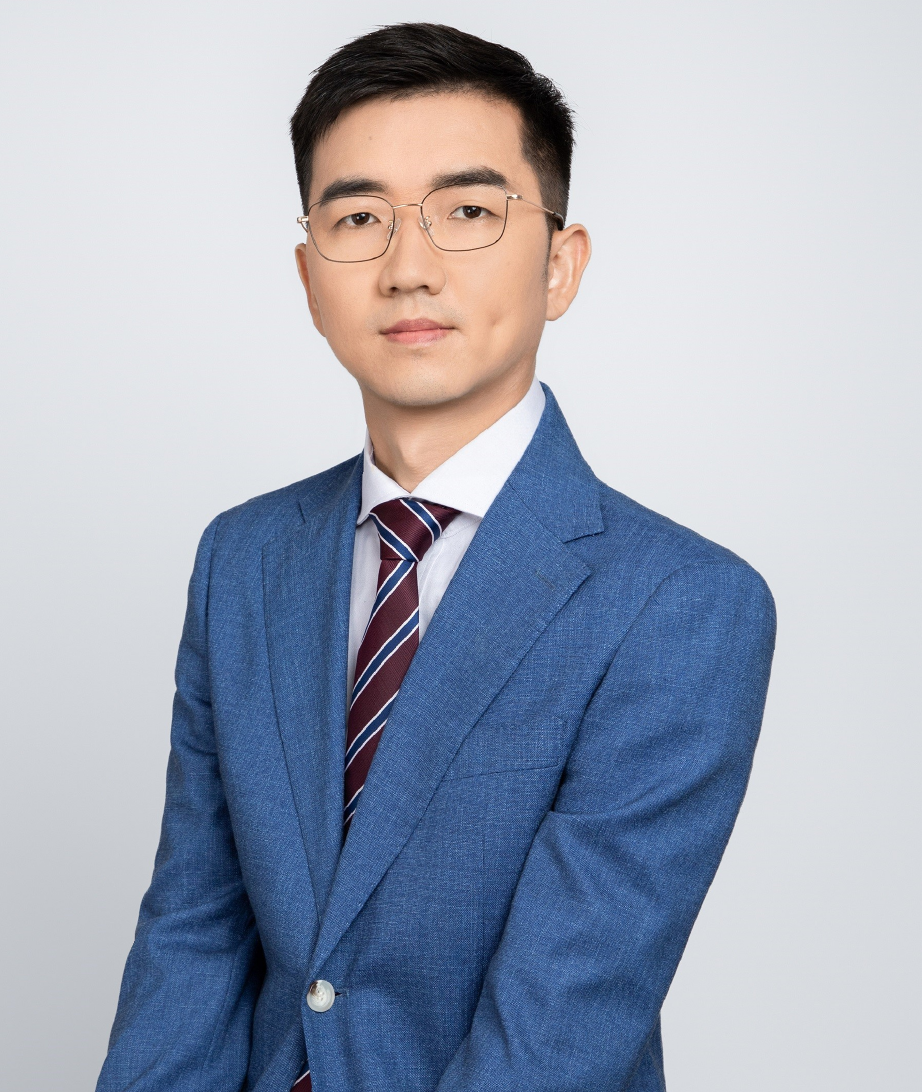}}]
{Ming He} received the PhD degree from USTC in 2018, and finished the post-doctor program at Shanghai Jiao Tong University in 2022. He is currently an advisory researcher of Lenovo Research, Beijing. He was a visiting scholar at UNC-Charlotte in 2016-2017; and was an algorithm researcher of DiDi, TAL in 2019-2021. His current research interests include recommender system, reinforcement learning, computer vision, and quantum machine learning. He has published 10+ papers in refereed journals and conference proceedings, e.g., WWW, SIGIR, TWEB, TIP, FCS, IJCAI and DASFAA. He received the KSEM 2018 Best Research Paper Award.
\end{IEEEbiography}
\vspace{-20pt}

\begin{IEEEbiography}[{\includegraphics[width=1in,height=1.25in,clip,keepaspectratio]{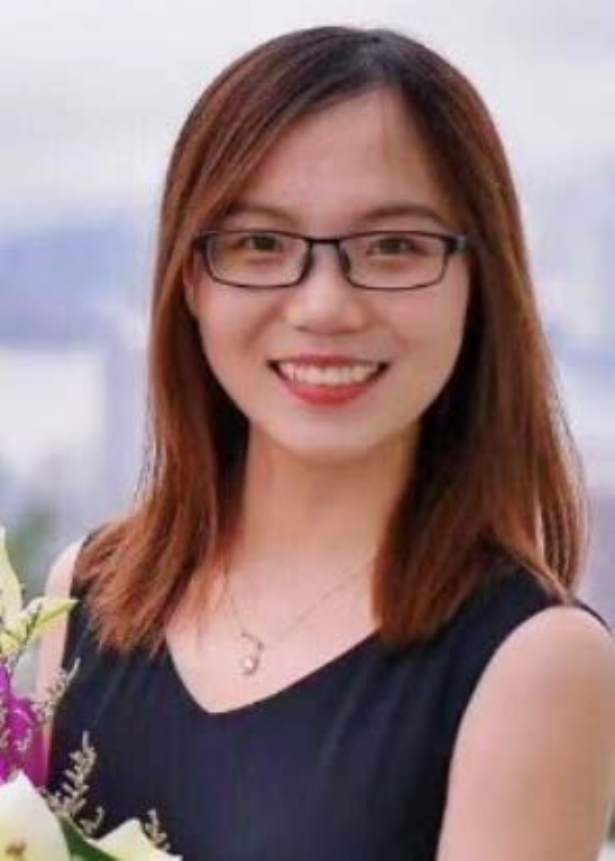}}]{Xiaomeng Li} (Membe, IEEE) received the Ph.D.
degree from The Chinese University of Hong Kong.
She is currently an Assistant Professor of electronic
and computer engineering at The Hong Kong University of Science and Technology. Her research lies
in the interdisciplinary areas of artificial intelligence
and medical image analysis, aiming at advancing
healthcare with machine intelligence.
\end{IEEEbiography}

\vspace{-20pt}

\begin{IEEEbiography}[{\includegraphics[width=1in,height=1.25in,clip,keepaspectratio]{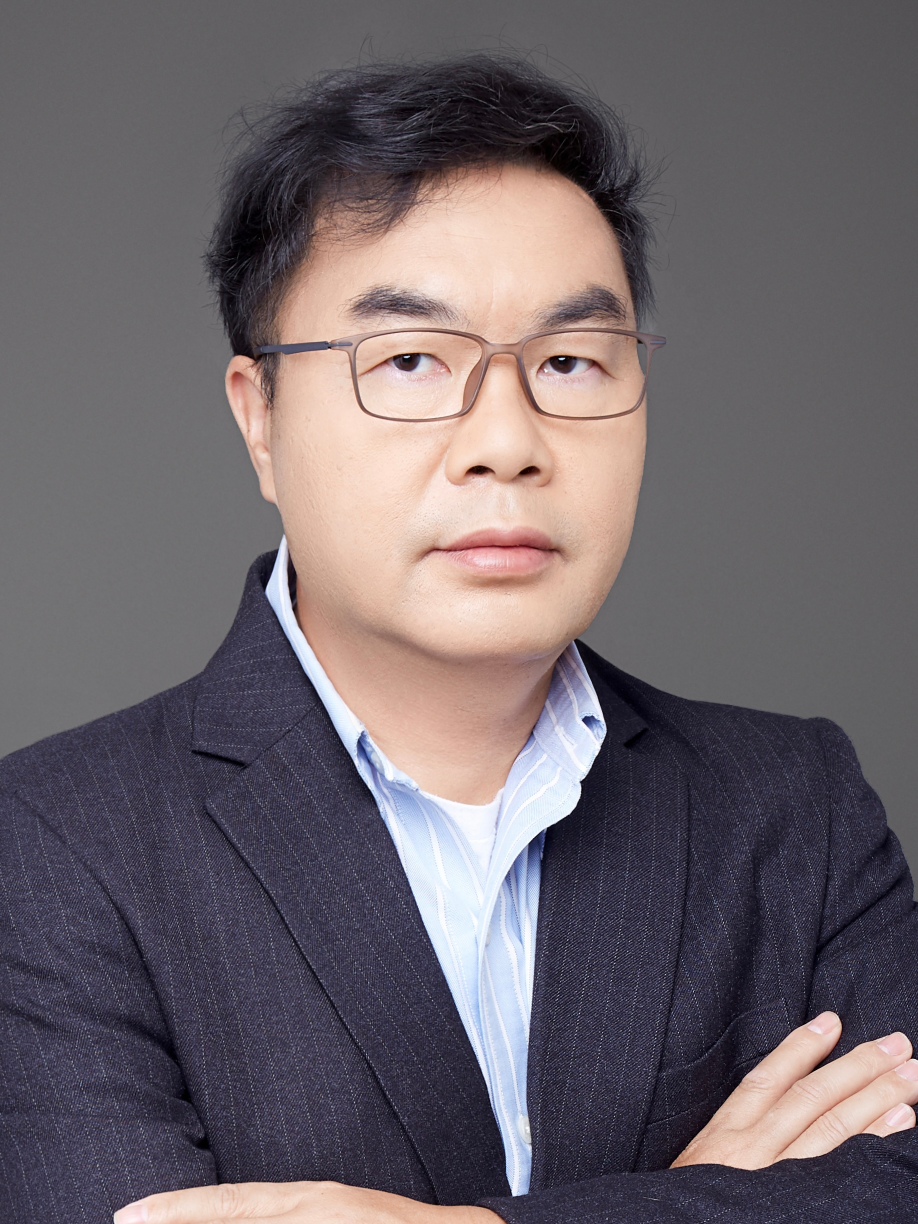}}]
{Jianping Fan} received the M.S. degree in theory physics from Northwest University, Xi’an, China, in 1994, and the Ph.D. degree in optical storage and computer science from the Shanghai Institute of Optics and Fine Mechanics, Chinese Academy of Sciences, Shanghai, China, in 1997. He was a Researcher with Fudan University, Shanghai, from 1997 to 1998. From 1998 to 1999, he was a Researcher with the Japan Society of Promotion of Science (JSPS), Osaka University, Japan. From 1999 to 2001, he was a Postdoctoral Researcher with the Department of Computer Science, Purdue University, West Lafayette, IN, USA. He is currently a leader of AI Lab of Lenovo Research. His research interests include large-scale deep learning, image/video privacy protection, automatic image/video understanding.
\end{IEEEbiography}

\vspace{-15pt}

\end{document}